\documentclass[%
 reprint,
superscriptaddress,
 amsmath,amssymb,
 aps,longbibliography,
]{revtex4-1}

\usepackage[colorlinks=true,linkcolor=blue,urlcolor=blue,citecolor=blue]{hyperref}
\usepackage{overpic}
\usepackage{physics}
\usepackage{graphicx}
\usepackage{dcolumn}
\usepackage{bm}
\usepackage{outlines}
\usepackage[dvipsnames]{xcolor}

\graphicspath{{./Figures/}} 

\definecolor{matlabblue}{rgb}{0,0.4470,0.7410}
\definecolor{matlabred}{rgb}{0.6350,0.0780,0.1840}
\definecolor{matlabpurple}{rgb}{0.4940,0.1840,0.5560}
\definecolor{matlaborange}{rgb}{0.8500,0.3250,0.0980}

\begin{document}




\title{Thermally-Polarized Solid-State Spin Sensor}

\author{Reginald Wilcox}
\affiliation{Massachusetts Institute of Technology, Cambridge, MA 02139, USA}
\affiliation{MIT Lincoln Laboratory, Lexington, MA 02421, USA}
\author{Erik Eisenach}%
\affiliation{Massachusetts Institute of Technology, Cambridge, MA 02139, USA}
\affiliation{MIT Lincoln Laboratory, Lexington, MA 02421, USA}
\author{John Barry}
 \email{john.barry@ll.mit.edu}
 \affiliation{MIT Lincoln Laboratory, Lexington, MA 02421, USA}
\author{Matthew Steinecker}
 \affiliation{MIT Lincoln Laboratory, Lexington, MA 02421, USA}
\author{Michael O'Keeffe}
 \affiliation{MIT Lincoln Laboratory, Lexington, MA 02421, USA}
\author{Dirk Englund}
\affiliation{Massachusetts Institute of Technology, Cambridge, MA 02139, USA}
\author{Danielle Braje}
 \affiliation{MIT Lincoln Laboratory, Lexington, MA 02421, USA}

\date{\today}

\begin{abstract}


Quantum sensors based on spin defect ensembles have seen rapid development in recent years, with a wide array of target applications.
Historically, these sensors have used optical methods to prepare or read out quantum states.
However, these methods are limited to optically-polarizable spin defects, and the spin ensemble size is typically limited by the available optical power or acceptable optical heat load.
We demonstrate a solid-state sensor employing a non-optical state preparation technique, which harnesses thermal population imbalances induced by the defect's zero-field splitting.
Readout is performed using the recently-demonstrated microwave cavity readout technique, resulting in a sensor architecture that is entirely non-optical and broadly applicable to all solid-state paramagnetic defects with a zero-field splitting.
The implementation in this work uses Cr$^{3+}$ defects in a sapphire (Al$_2$O$_3$) crystal and a simple microwave architecture where the host crystal also serves as the high quality-factor resonator.
This approach yields a near-unity filling factor and high single-spin-photon coupling, producing a magnetometer with a broadband sensitivity of 9.7~pT/$\sqrt{\mathrm{Hz}}$.

\vspace{5mm}

\end{abstract}

\maketitle

\UseRawInputEncoding
\section{Introduction}

\begin{figure*}[t] 
\hspace{-2mm}
\begin{minipage}[b]{0.3\textwidth}
\begin{overpic}[width=2.1in]{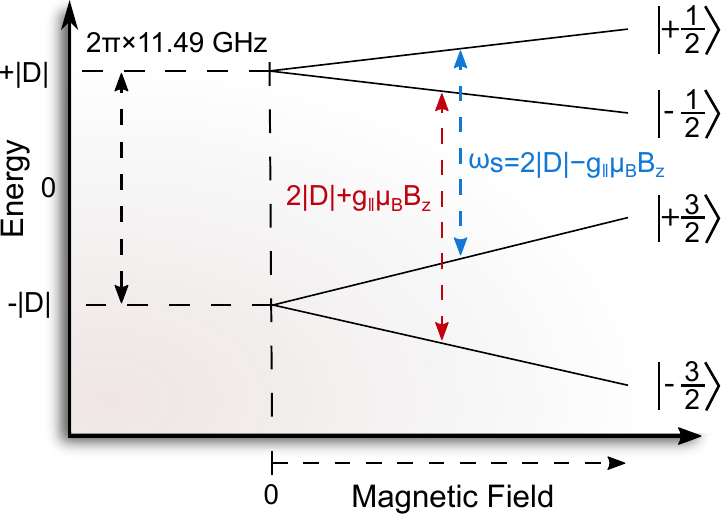} \put(-10,75){\textbf{a)}}
\end{overpic}
\end{minipage}
\;
\begin{minipage}[b]{0.3\textwidth}
\begin{Overpic}{\put(-72,20){\includegraphics[width=2.1in]{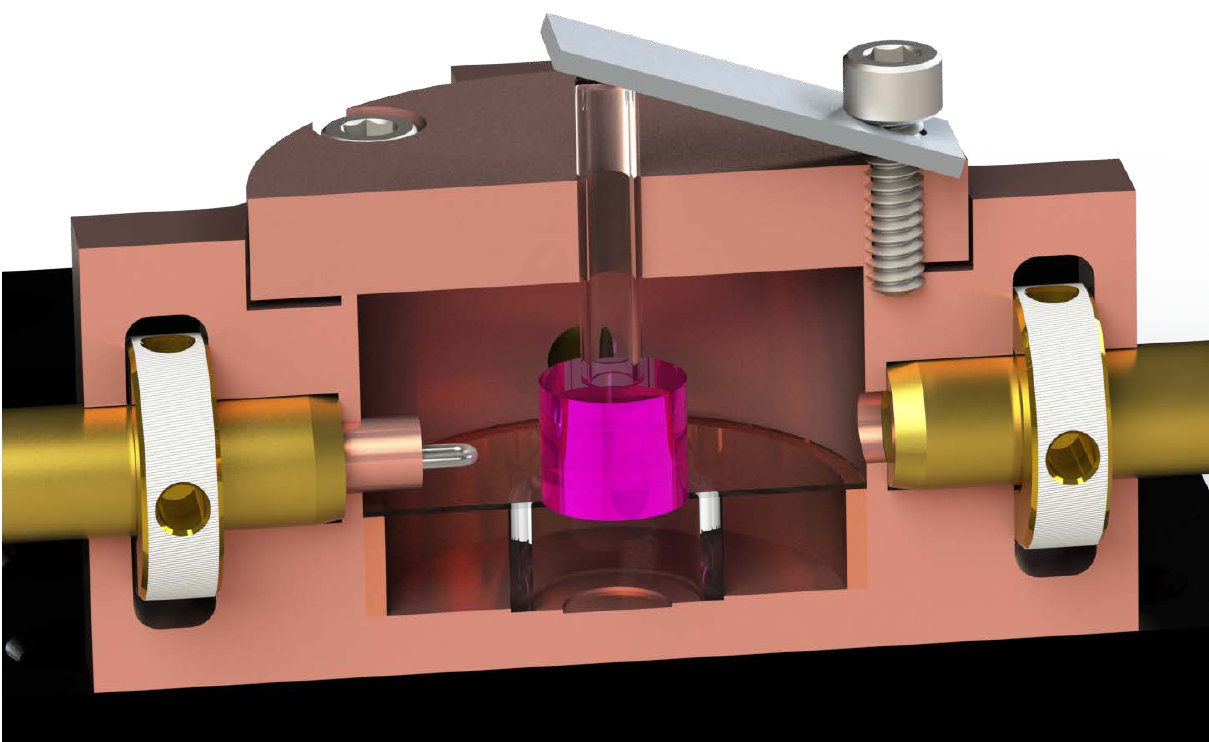}}} \put(-72,100){\textbf{b)}} 
\end{Overpic}
\end{minipage}
\;
\begin{minipage}[b]{0.3\textwidth}
\begin{overpic}[width=2.1in]{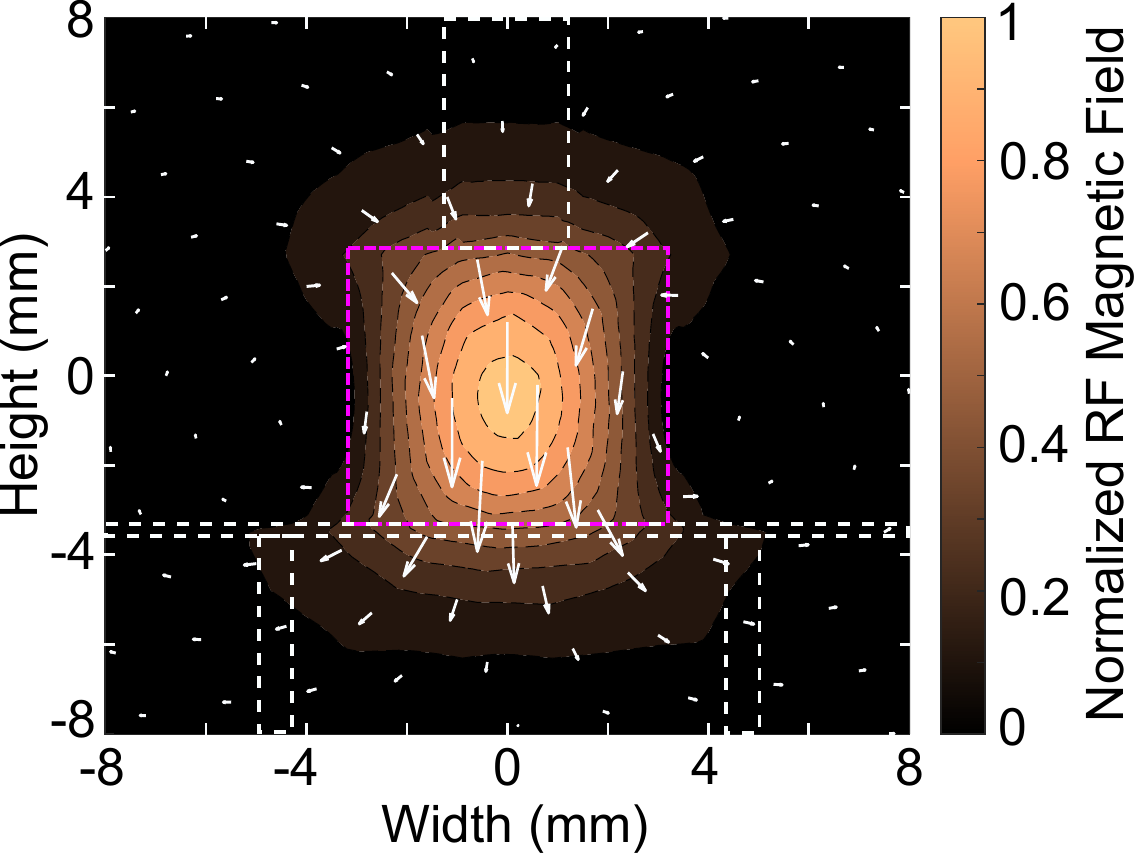} \put(-5,75){\textbf{c)}} 
\end{overpic}
\end{minipage}
\caption{\textbf{Ruby resonator design.} \textbf{a)} The electronic ground state of trivalent chromium (Cr$^{3+}$) in sapphire consists of two pairs of degenerate spin doublets separated by a $2\pi \times 11.49$~GHz ZFS. 
External magnetic fields applied along the crystallographic c-axis lift the degeneracy of the $|m_s = \pm \tfrac{1}{2}\rangle$ and $|m_s = \pm \tfrac{3}{2}\rangle$ sublevels as shown, allowing the $|m_s = + \tfrac{3}{2}\rangle$ $\leftrightarrow$ $|m_s = + \tfrac{1}{2}\rangle$ transition with frequency $\omega_s$ to be addressed spectroscopically.
\textbf{b)} Device cross section. MWs are coupled into the ruby resonator shown in pink, with a single probe loop on the left.
While the shield supports up to four probe loops, only a single probe is used in this work.
The ruby resonator is centered on a semi-insulating wafer of silicon carbide (SiC) and is clamped in place by a spring-loaded fused-silica tube.
A copper shield encloses the ruby resonator and decreases radiative losses from decreasing the resonator's quality-factor.
\textbf{c)} Simulated RF magnetic field distribution of the ruby resonator's $\text{TE}_{01\delta}$ mode.
The RF magnetic field is denoted by white arrows, the ruby resonator is outlined in pink, and the SiC and fused-silica support structures are indicated by white dashed lines. }
\label{fig:cavity}
\end{figure*}



Solid-state spin systems have transitioned from physics demonstrations to promising quantum sensors \cite{karadas2018feasibility,sturner2021integrated, barry2016optical,davis2018mapping,arai2021millimetrescale,Webb2021detection,fescenko2019diamond,shi2015single,zhou2021imaging,turner2020magnetic,patel2020subnanotesla,bertelli2020magnetic,lenz2021imaging,jenkins2020imaging,hsieh2019imaging,lovchinsky2016nuclear} as performance has improved through materials engineering \cite{achard2020chemical,tallaire2020high,alsid2019photoluminescence,ashfold2020nitrogen,bauch2020decoherence,wolfowicz2021quantum,edmonds2021characterisation,watanabe2021creation,bluvstein2019identifying}, coherent control \cite{Aslam2017nuclear,waeber2019pulse,o2019hamiltonian,glenn2018high,bauch2018ultralong,smits2019two,bargill2013solid,dreau2011avoiding,hart2021diamond,pham2012enhanced}, and novel readout \cite{neumann2010,shields2015,eisenach2021cavity,ebel2021dispersive,bourgeois2015photoelectric,hopper2016near,hopper2018spin,niethammer2019}.
While performance now rivals atomic-based sensors \cite{barry2020sensitivity,kehayias2017solution,fescenko2020diamond,bucher2020hyperpolarization,fu2020sensitive,cappellaro2017quantumsensing}, solid-state systems still maintain much of the complexity of their atomic counterparts \cite{cappellaro2017quantumsensing}.
Whether measuring electromagnetic fields, temperature, or time \cite{taylor2008high,kucsko2013nanometre,neumann2013high,zhang2021robust,nishimura2021widefield,michl2019robust,yang2020vector,chen2017high,cappellaro2017quantumsensing}, spin sensors perform three primary processes \cite{barry2020sensitivity, cappellaro2017quantumsensing}: quantum state initialization, interaction with the environment, and readout.
Initialization has seen little recent development and is nearly universally achieved optically \cite{wolfowicz2021quantum}.
This limits sensing species to optically-polarizable defects, with the bulk of experimental effort dedicated to nitrogen-vacancy (NV) diamond \cite{achard2020chemical}.
Moreover, the light source is often the primary driver of device complexity and power consumption.
To overcome these limitations, we introduce a novel quantum sensor using chromium ions (Cr$^{3+}$) in sapphire (also known as ruby).


The key advance of this work is combining initialization and readout techniques to enable a simple sensor architecture which maintains performance.
In particular, this sensor harnesses the crystal zero-field splitting (ZFS) to thermally polarize the spins at room temperature and extends the recently-demonstrated microwave (MW) cavity readout technique \cite{eisenach2021cavity, ebel2021dispersive} to a non-NV-diamond system for the first time.
Together, these advances result in an entirely MW-based (non-optical) device that exhibits high sensitivities when used as a magnetometer.
More broadly, this work serves as proof of concept for a wide variety of sensing modalities employing diverse solid-state defects, which need not be optically-polarizable.

\section{Thermal Spin Polarization in Ruby}
In contrast to optically-polarized sensors, the device demonstrated here employs passive thermal spin polarization for quantum state initialization.
Thermal spin polarization relies on energy differences between quantum states to induce population differences via the Boltzmann distribution.
In related systems, such as those used in EPR spectroscopy or in experiments probing cavity QED effects, this energy difference is typically produced by strong magnetic fields \cite{poole1996electron,loubser1978electron,eaton2010quantitative}; the population difference can then be enhanced by operating at cryogenic temperatures \cite{le2016towards,angerer2017ultralong,probst2013anisotropic}.
For example, a 0.41~T field (corresponding to a $2\pi\times11.5$~GHz transition) results in 0.094\% polarization for a two-level electronic system at room temperature \cite{cammack2013}.
However, certain crystals exhibit an electric field anisotropy which induces a ZFS, providing an energy difference between spin states without the need for large magnetic fields.
In the case of the Cr$^{3+}$ ions in ruby with spin $S=\tfrac{3}{2}$, a net population difference between the $|m_s = +\tfrac{3}{2}\rangle$ and $|m_s = +\tfrac{1}{2}\rangle$ states is induced by the crystal ZFS.
At 293~K, the net population difference between these two states is 0.047\% of the total number of Cr$^{3+}$ ions, corresponding to a polarized spin density of $\sim 7\times10^{15}$~cm$^{-3}$ (see Supplemental Sec.~\ref{supp:thermal}).

Ruby has several key properties which make it well suited as a host crystal for MW cavity readout.
First, the ruby crystal can serve as the dielectric resonator, which vastly improves the filling factor over volume-limited crystals such as nitrogen-vacancy diamond coupled to an external resonator.
Second, the ruby resonator supports a high quality-factor mode ($\sim 50,000$) near the spin transition frequency  \cite{blair1982} enabling a large single-spin-photon coupling.
Third, ruby crystal growth is a mature technology, with high-purity single-crystal material available in liter-scale volumes \cite{sapphiregrowth2003, kurlov2016}.
Finally, ruby's anisotropy induces sufficient thermal spin polarization for high-sensitivity magnetometry at room temperature.
Together, these properties allow creation of a quantum sensor free of lasers, large magnetic fields, and cryogenics.
Because the sensing head has a volume less than 100~cm$^3$ and the supporting RF and readout electronics are commercially available, this demonstration offers promise for near-term scalable production of solid-state quantum sensors with academic, commercial, and industrial applications.

\section{Theoretical Background}
Operation of ruby as a magnetometer is explained by its Hamiltonian under application of an external magnetic field $\mathbf{B} = \left[B_x,B_y,B_z\right]$.
With the $z$-axis defined parallel to the ruby c-axis, the Cr$^{3+}$ ground-state Hamiltonian is \cite{rubynist1977}
\begin{eqnarray}
\mathcal{H} &=& g_{\parallel}\mu_B B_z S_z + g_{\perp}\mu_B(S_x B_x + S_y B_y) \label{eqn:hamiltonian}  \\
&&+ D\left[S_z^2 - \tfrac{1}{3}S(S + 1)\right], \nonumber
\end{eqnarray}
\noindent
where $\mu_B$ is the Bohr magneton, $g_{||}$ and $g_{\perp}$ are the axial and transverse g-factors, and $D$ is the ZFS parameter.
The total spin angular momentum is $S = \tfrac{3}{2}\hbar$, and $S_x$, $S_y$, and $S_z$ are the spin matrices.
At room temperature, the ZFS is $2D \approx - 2\pi \times 11.49$~GHz, and $g_{\parallel} \approx g_{\perp} \approx 2$ \cite{rubynist1977}.
All Cr$^{3+}$ sites are magnetically equivalent.

Without an applied magnetic field, the eigenstates form a Kramers doublet with degenerate pairs $|m_s = \pm\tfrac{1}{2}\rangle$ and $|m_s = \pm\tfrac{3}{2}\rangle$ \cite{sewani2020}.
Both degeneracies are lifted in the presence of an external field through the Zeeman effect.
For magnetic fields applied parallel to the c-axis, the eigenenergies vary linearly with $S_z B_z$, as shown in Fig.~\ref{fig:cavity}a~\cite{rubynist1977}.
In this work, we primarily consider magnetic fields parallel to the c-axis (see Supplemental Sec.~\ref{supp:hamiltonian}).
If the eigenenergies are accurately measured, their magnetic field dependence can be exploited for high-sensitivity magnetometry.

Microwave cavity readout interrogates the spin transition frequency $\omega_s$ by applying a MW probe signal at frequency $\omega_d$ to the ruby resonator and measuring the reflected signal \cite{eisenach2021cavity}.
Shifts in $\omega_s$ are encoded in the spin ensemble's absorptive and dispersive modification of the reflection coefficient $\Gamma$.
The complex-valued voltage reflection coefficient is~\cite{eisenach2021cavity}
\begin{equation}\label{eqn:reflection}
\Gamma = -1 + \frac{\kappa_{c1}}{\frac{\kappa_c}{2} + i(\omega_d - \omega_c) + \Pi},
\end{equation}
where $\kappa_{c} \equiv \kappa_{c0} + \kappa_{c1}$ is the loaded cavity linewidth, $\kappa_{c0}$ and $\kappa_{c1}$ are the intrinsic cavity linewidth and input couplings rates, $\omega_c$ is the bare cavity resonance, and $\Pi$ is an interaction term incorporating the absorptive and dispersive effects of the spins on the ruby resonator. The interaction term $\Pi$ may be written as \cite{eisenach2021cavity}
\begin{equation} \label{eqn:spininteraction}
    \Pi = \frac{g_s^2 N}{\frac{\kappa_s}{2} + i(\omega_d - \omega_s) + \frac{g_s^2 n_{\mathrm{cav}}\kappa_s/(2\kappa_{\mathrm{th}})}{\frac{\kappa_s}{2} - i(\omega_d - \omega_s)}},
\end{equation}
\noindent
where $\kappa_s = 2/T_2$ is the homogeneous width of the spin resonance, $\kappa_{\mathrm{th}} = 1/T_1$ is the thermal polarization rate, $n_{\mathrm{cav}}$ is the average number of MW photons in the cavity over the cavity lifetime 1/$\kappa_c$, and $N$ is effective number of polarized spins, given by the population difference between the $|m_s = +\tfrac{3}{2}\rangle$ and $|m_s = +\tfrac{1}{2}\rangle$ states.
The single spin-photon coupling is $g_s = \tfrac{\gamma n_{\perp}}{2}\sqrt{\tfrac{\hbar\omega_c \mu_0}{V_{\mathrm{cav}}}}$, where $\gamma$ is the electron gyromagnetic ratio, $V_{\mathrm{cav}}$ is the ruby resonator modal field volume, $\mu_0$ is the vacuum permeability, and $0\leq n_{\perp} \leq 1$ is a geometric factor, which occurs because only fields transverse to the spin quantization axis can drive transitions.
The reflection coefficient $\Gamma$ is maximally sensitive to changes in magnetic field when all frequencies are equal: $\omega_c = \omega_d = \omega_s$.
By measuring the reflection coefficient for various values of $\omega_s$, $\omega_d$, and applied MW power, the spin-cavity interaction can be characterized.
Such characterization allows straightforward optimization of the device as a magnetometer.

\section{Experimental Setup}

\subsection{Microwave Cavity}
Previous room-temperature spin-cavity coupling experiments employed a solid-state spin system coupled to an external MW cavity \cite{eisenach2021cavity,breeze2017room,breeze2018continuous,ebel2021dispersive}.
In this work, the MW cavity is formed by the host crystal of the spin system itself.
A monocrystalline ruby sample with a Cr$^{3+}$ concentration of 0.05\% by weight (see Supplemental Sec.~\ref{supp:thermal}) is used to produce a TE$_{01\delta}$ cavity mode near the Cr$^{3+}$ spin resonance.
The resulting ruby resonator is a cylinder 5.85~mm tall and 6.98~mm in diameter \cite{jbreeze2016}, with the c-axis oriented radially.
With the given dimensions and relative dielectric constants of $\epsilon_{||} = 11.5$ and $\epsilon_{\perp} = 9.4$ (parallel and perpendicular to the c-axis, respectively) \cite{krupka1999, kobayashi1993, jbreeze2016}, the ruby resonator exhibits a TE$_{01\delta}$ resonance (Fig \ref{fig:cavity}c) at $\omega_c \approx 2\pi\times 11.4$~GHz with a filling factor $\zeta\approx 0.7$ (see Supplemental Sec.~\ref{supp:ff}).
As constructed, the ruby resonator has an unloaded quality factor of $Q_0 = 35,000$
\footnote{Without the SiC, which lowers the quality-factor due to a higher loss tangent \cite{parshin2017,krupka1999} and spreading of the mode, we measured $Q_0 = 50,000$.}.

The resonator is centered coaxially in a cylindrical copper shield with inner height of 16.1~mm and inner diameter of 24~mm \cite{jbreeze2016}, as shown in Fig.~\ref{fig:cavity}b.
This thin shield (5~mm) improves the quality factor by mitigating radiative losses at 11.4~GHz, but does not significantly disturb DC and low frequency magnetic fields of interest.
The ruby resonator is supported mechanically by a 330~$\mu\mathrm{m}$ thick silicon carbide (SiC) wafer and a spring-loaded fused silica tube.
By providing a strong thermal link to the copper shield, the SiC reduces temperature fluctuations due to the MW-induced heat load on the crystal, thereby decreasing temperature-dependent shifts in the cavity resonance \cite{tobar1996sapphireresonator}.


The shielded ruby resonator is placed in a uniform magnetic field $\mathbf{B_0}$ created by an electromagnet, which is driven by a bipolar operational amplifier power supply.
The field is oriented along the c-axis of the ruby resonator.
This bias magnetic field lifts the degeneracy of the  $|\pm \tfrac{1}{2}\rangle$ and $|\pm \tfrac{3}{2}\rangle$ spin states, allowing the $|+\tfrac{1}{2}\rangle \leftrightarrow |+\tfrac{3}{2}\rangle$ transition (with angular frequency $\omega_s$) to be addressed spectroscopically as an effective two-level system \footnote{{The spin resonance has an observed linewidth  $\kappa_s = 2\pi\times 42$~MHz, corresponding to 15~G. Thus, for bias magnetic fields $\gtrsim 15$~G, the $|+\tfrac{1}{2}\rangle \leftrightarrow |+\tfrac{3}{2}\rangle$ transition is spectrally separated from all other transitions.}}.
For magnetometry data in this work, $B_0 \approx 31$~G is chosen so that the spin resonance at $\omega_s$ and cavity resonance at $\omega_c$ are nearly equal. 
The probe MWs, at angular frequency $\omega_d$, are then tuned so $\omega_d \approx \omega_c \approx \omega_s$ to maximize the spin-cavity interaction.


\subsection{Microwave Signal Chain}
\begin{figure}[t]
\hspace{-2mm}
\includegraphics[width=3.3in]{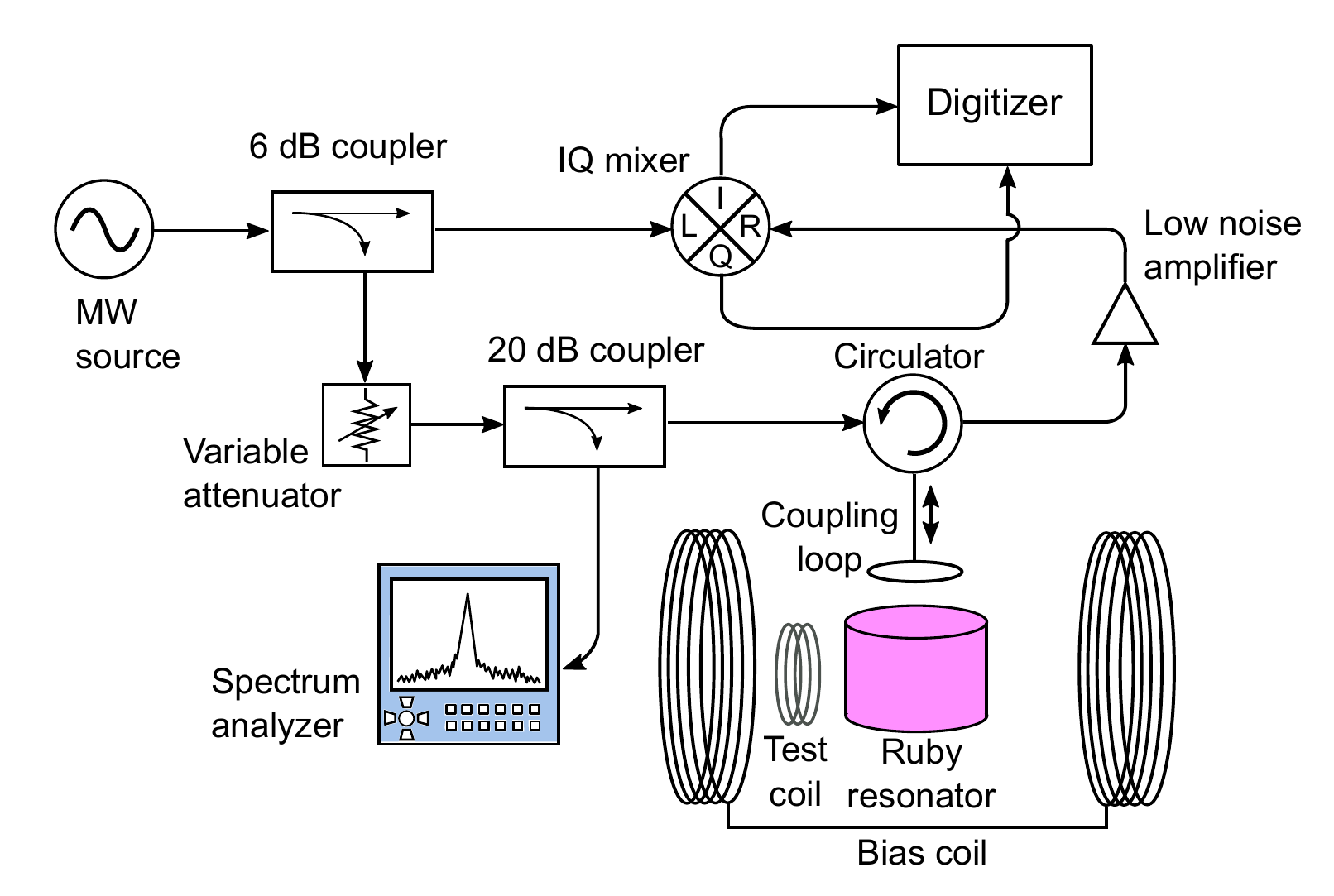}
\caption{\textbf{Experimental setup and MW signal chain.} A probe MW signal near-resonant with the ruby resonator is split using a 6 dB directional coupler into a signal arm and a reference arm. 
Incident microwaves in the signal arm probe the ruby resonator through a circulator.
MW signals exiting the circulator are then amplified by a low-noise amplifier (LNA) and mixed to baseband using an IQ mixer. The spectrum analyzer monitors the MW power incident on the ruby resonator.}\label{fig:deviceschematic}
\end{figure}

In the MW cavity readout technique, the spin resonance frequency $\omega_s$ is inferred from the ensemble's dispersive shift of the cavity resonance~\cite{eisenach2021cavity,ebel2021dispersive}.
The implementation in this work interrogates the cavity resonance using a homodyne receiver structure, as shown in Fig.~\ref{fig:deviceschematic}.
The probe MWs are first split into a signal arm and a reference arm.
Next, the signal arm is attenuated as needed, sampled by a spectrum analyzer through a 20-dB directional coupler, and delivered to a circulator.
Microwaves out of the circulator are coupled into the ruby resonator using a probe loop inserted into the copper shield.
By using a circulator, incident MWs are isolated from those reflected by the resonator, allowing the complex reflection coefficient $\Gamma$ to be measured.
A 24-dB-gain low-noise amplifier (LNA) amplifies the reflected MWs, which are terminated in an IQ mixer's RF port.
The IQ mixer's LO port is driven by the reference arm.

The signal is downconverted to baseband by the IQ mixer, and both the I and Q channels are digitized.
Capturing both the I and Q mixer channels allows the reflected signal to be reconstructed in software as if there were a physical phase shifter (see Supplemental Sec.~\ref{supp:iqmixing}).
The software reconstruction is configured to separate the signal into an absorptive channel and a dispersive channel.

\section{Experimental Results}

\subsection{Spin-cavity interaction}
\label{sec:spininteraction}
\begin{figure}[ht]
\centering
\includegraphics[width=3.3in]{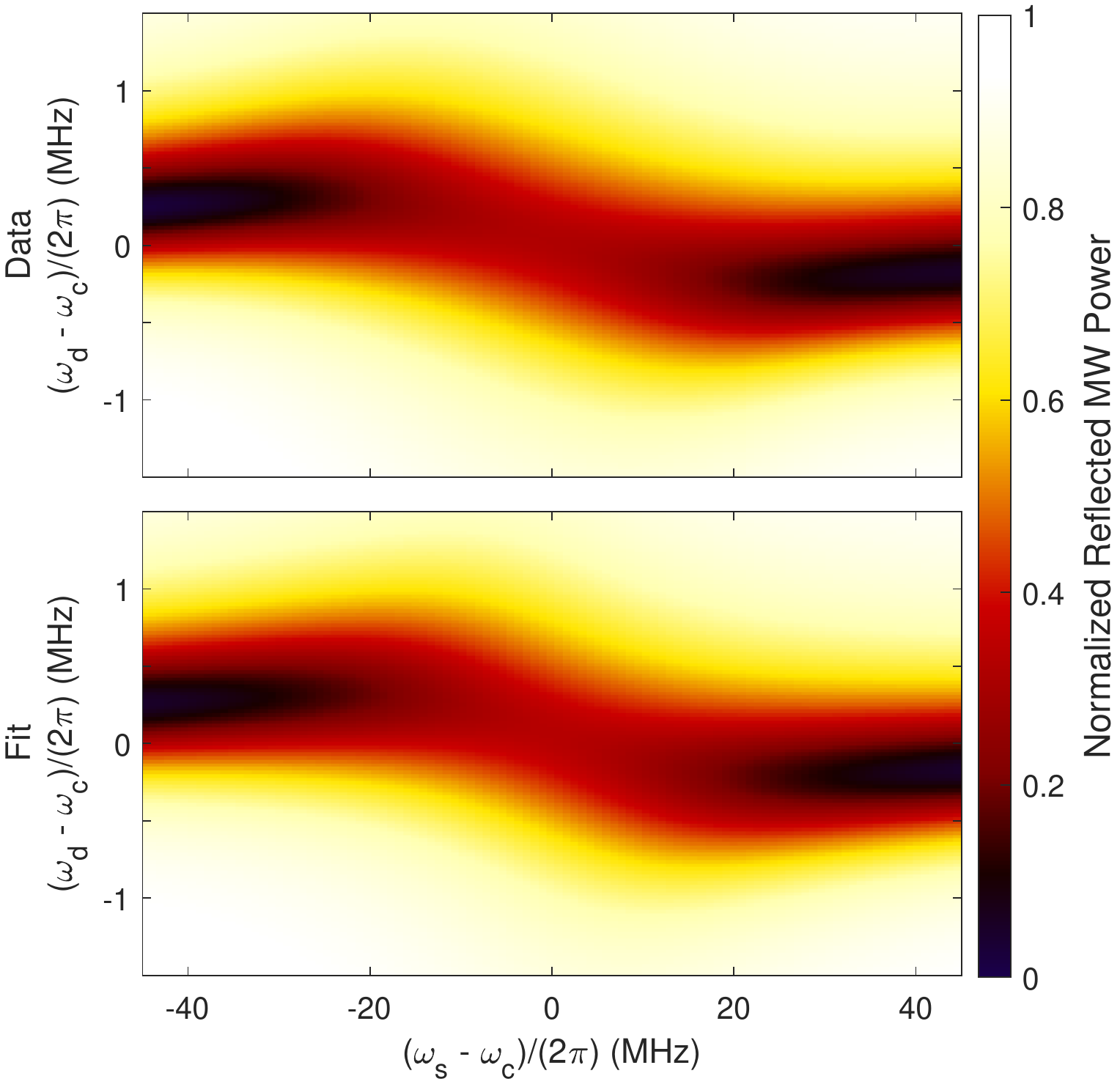}
\put(-240,218){\textbf{a)}}
\put(-240,110){\textbf{b)}}
\caption{\textbf{Ensemble-cavity coupling.}
a) The spin resonance frequency is swept across the bare cavity resonance (horizontal axis) by varying the bias magnetic field; by concurrently varying the MW drive frequency (vertical axis), an avoided crossing due to the ensemble-resonator coupling is observed.
b) Simulation accurately reproduces experimental results, providing best-fit system parameters as described in the main text.
Data were recorded with 0~dBm applied to the ruby resonator under critical coupling, and the reflection coefficient is normalized to unity.
}
\label{fig:avoidedcrossing}
\end{figure}

To quantify the strength of the interaction between the cavity and spin-ensemble, a low-power MW signal is applied to the ruby resonator.
The reflection coefficient is measured as the MW drive frequency $\omega_d$ and the magnetic bias field (and thus the spin resonance frequency $\omega_s$) are independently varied \footnote{The LNA is removed from the signal chain during this process to reduce amplitude-dependent distortion on the reflected signal.}.
Figure~\ref{fig:avoidedcrossing} depicts the observed and simulated avoided crossing with 0 dBm applied to the ruby resonator.
The best fit parameters derived from Equations~\eqref{eqn:reflection} and \eqref{eqn:spininteraction} are $N = 3.5\times 10^{14}$ polarized Cr$^{3+}$ spins, $g_\text{eff} \equiv g_s \sqrt{N} = 2\pi \times 3.5$~MHz, $\kappa_c = 2\pi \times 660$~kHz, $\kappa_s = 2\pi \times 42$~MHz, and $\kappa_{\mathrm{th}} = 2\pi\times 120$~kHz (see Supplemental Sec.~\ref{supp:parameterestimation}).
Given the Cr$^{3+}$ concentration and the modal field volume, we compute $N_{\mathrm{tot}} = 8\times 10^{17}$ Cr$^{3+}$ spins being interrogated.
This corresponds to an effective polarization of $\mathcal{P} = 0.044\%$, in agreement with the expected value (see Supplemental Sec.~\ref{supp:thermal}).
The collective cooperativity, defined as $\xi = 4 g^2_\text{eff} / (\kappa_s \kappa_c)$, is a dimensionless figure of merit describing the strength of the ensemble-cavity interaction \cite{tanjisuzuki2011}.
The given fit parameters result in a collective cooperativity of $\xi = 1.8$, indicating operation in the high-cooperativity regime ($\xi > 1$).

\begin{figure}[t]
\begin{minipage}[b]{0.23\textwidth}
\begin{overpic}[width=1.7in]{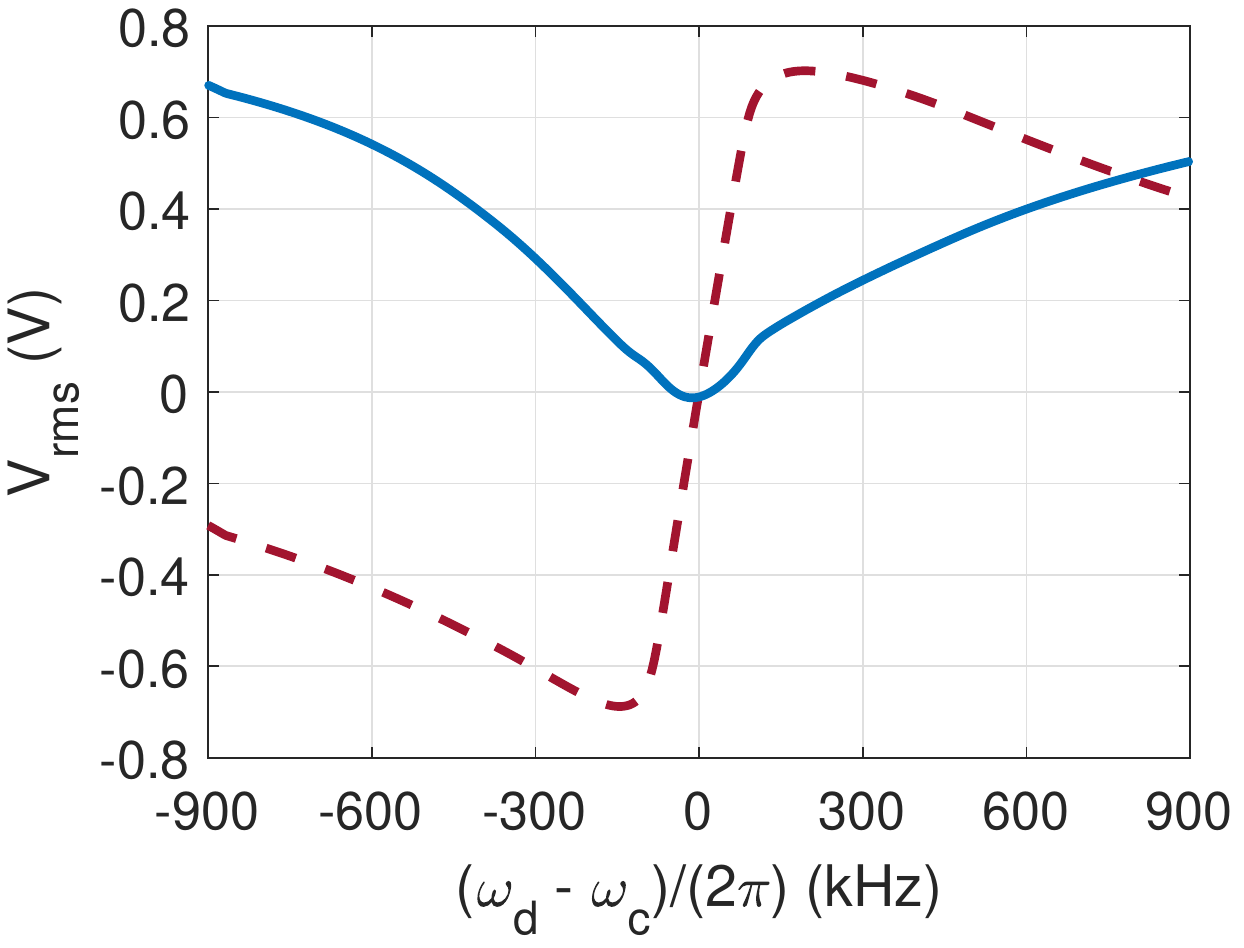} \put(0,75){\textbf{a)}}
\end{overpic}
\end{minipage}%
\;
\begin{minipage}[b]{0.23\textwidth}
\begin{overpic}[width=1.7in]{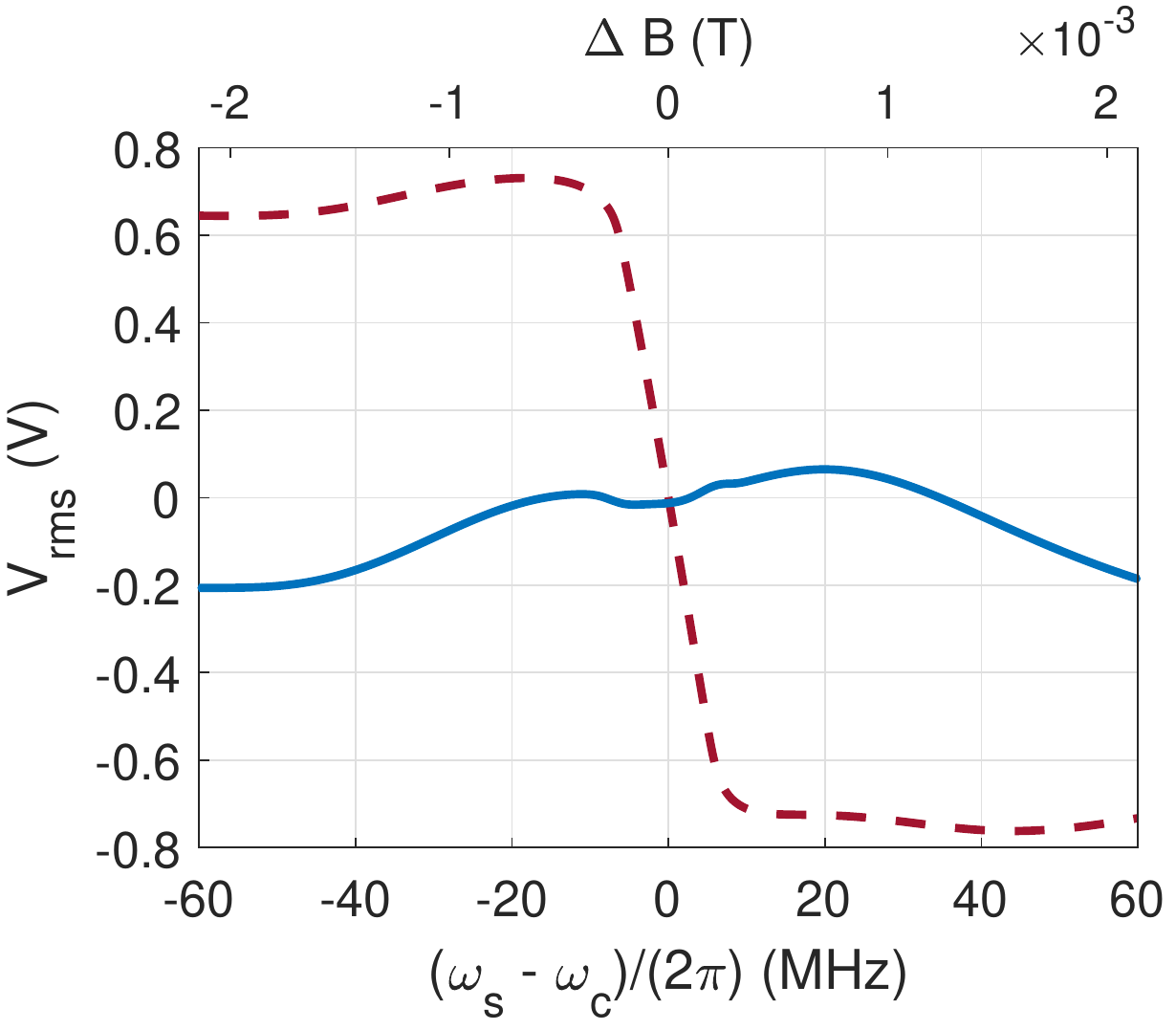} \put(0,75){\textbf{b)}}
\end{overpic}
\end{minipage}
\caption{\textbf{Reflection measurements of the spin-cavity response.}  \textbf{a)} Reflected absorptive (\textcolor{matlabblue}{\rule[0.75mm]{3mm}{.25mm}}) and dispersive (\textcolor{matlabred}{\rule[0.75mm]{1.1mm}{.25mm}~\rule[0.75mm]{1.1mm}{.25mm}}) voltage signal as a function of drive-cavity detuning $(\omega_d-\omega_c)$. The dispersive shift is near-linear around the zero-crossing. The asymmetry in the cavity reflection profile is due to non-monotonic amplifier gain near saturation. \textbf{b)} Reflected absorptive (\textcolor{matlabblue}{\rule[0.75mm]{3mm}{.25mm}}) and dispersive (\textcolor{matlabred}{\rule[0.75mm]{1.1mm}{.25mm}~\rule[0.75mm]{1.1mm}{.25mm}}) voltage signal as a function of spin-cavity detuning $(\omega_s-\omega_c)$. The dispersive shift varies approximately linearly with small changes in the spin-cavity detuning about zero.
The maximum slope on the dispersive component, $M_{\mathrm{max}} \approx 3000$~V/T, sets the device's response when operated as a magnetometer.
The slight observed asymmetry in both components is hypothesized to result from the off-resonant spin transition interacting with the ruby resonator mode and imperfect alignment of the bias field with the optical axis ($\lesssim 10^{\circ}$).
The applied MW power for both panels is 11~dBm.
}
\label{fig:cavitysweeps}
\end{figure}

To characterize the system's performance as a magnetometer, the system is probed under optimal applied MW power (see Supplemental Sec.~\ref{supp:power}).
The cavity interaction with probe MWs is observed by fixing the bias magnetic field so that $\omega_s\approx\omega_c$ then sweeping the MW drive frequency $\omega_d$ and monitoring the reflected signal, seen in Fig.~\ref{fig:cavitysweeps}a.
This is used to properly center the drive frequency for magnetometry and reveals the cavity linewidth $\kappa_c$, which is monitored to adjust the probe loop for critical coupling.
The dispersive and absorptive responses of the spin resonance are observed by fixing the MW drive frequency so that $\omega_d \approx \omega_c$ and sweeping the bias magnetic field field, seen in Fig.~\ref{fig:cavitysweeps}b.
(Iterating between sweeps of $\omega_d$ and $\omega_s$ allows these parameters to be accurately set to $\omega_c$.)
When the spin transitions are are resonant with the cavity, the dispersive component exhibits a sharp slope with maximum value $M_{\mathrm{max}} \approx 3000$~V/T.
This regime is of particular interest to magnetometry as the reflected MW signal is maximally sensitive to changes in the magnetic field \cite{eisenach2021cavity}.

\subsection{Magnetometry}
\label{sec:magnetometry}
Small changes in the external magnetic field may be detected by monitoring the dispersive shift of the reflected MW probe signal.
For high sensitivity magnetometry, the bias magnetic field is chosen so that the device operates in the region of maximal slope $M_{\mathrm{max}}$ where the reflected signal is most sensitive to the applied magnetic field.
Magnetometry measurements are performed using an independently applied AC test magnetic field. 
The chosen field magnitude is small compared to the spin resonance linewidth $\kappa_s$ so as not to perturb the system from the optimal configuration.
A low-noise signal generator drives a coil to generate a test field with amplitude $B^\text{RMS}_\text{test} = 242$~nT, frequency $\omega_m = 2\pi \times 10$~Hz, and orientation along the ruby c-axis.
This frequency is chosen high enough that flicker noise does not limit the digitization fidelity, but low enough that the test field is not significantly attenuated by the copper shield.
The amplitude of the applied test field is verified using three independent methods (see Supplemental Sec.~\ref{supp:testfieldamplitude}).

By computing the spectrum of the RMS voltage in the dispersive channel and evaluating the peak at the test field frequency $\omega_m$, which we denote $V_\text{m}$, we can calculate the projected magnetic sensitivity as
\begin{equation}
\label{eqn:sensitivity}
\eta = \frac{e_n}{V_\text{m}/B^\text{RMS}_\text{test}},
\end{equation}
\noindent
where $e_n$ is the RMS voltage noise floor of the single-sided spectrum and $B^\text{RMS}_\text{test}$ is the RMS amplitude of the applied test field.
There are competing effects which dictate the optimal interrogation MW power.
Increasing the MW power increases the signal amplitude, but also causes power broadening, decreasing the fractional change in signal for a given change in applied magnetic field \cite{abragam1983principles}.
Measurements suggest that 11~dBm of MW power and a bias field of $B_0 = 31$~G are optimal for the present system (see Supplemental Sec.~\ref{supp:power}).
This bias field $B_0$ produces $\omega_s \approx \omega_c$, consistent with the expectation that $\omega_s\approx \omega_c \approx \omega_d$ is optimal for magnetometry~\cite{rubynist1977,eisenach2021cavity}.

\begin{figure}[t]
\centering
\includegraphics[width=3.375in]{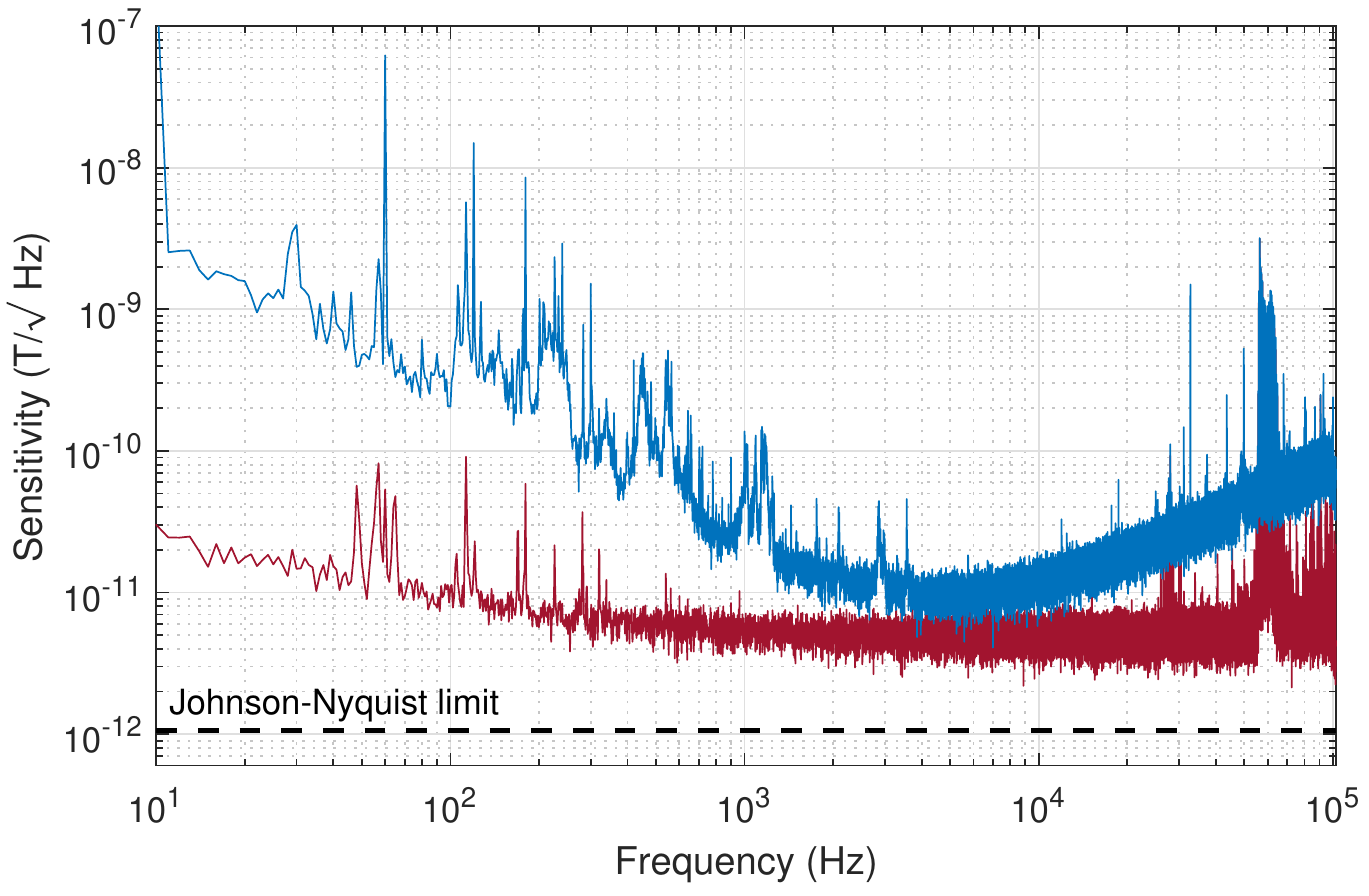}
\caption{\textbf{Broadband magnetometry using Cr$^{3+}$ ions in ruby.}
Magnetometry is performed by measuring MWs reflected off the ruby resonator using an IQ mixer. Based on the noise spectral density measured during magnetometer operation (\textcolor{matlabblue}{\rule[0.75mm]{3mm}{.25mm}}) we project a sensitivity of $\approx$10~$\mathrm{pT}\sqrt{\text{Hz}}$ in the low-noise band between 4~kHz and 6~kHz.
The projected sensitivity in the low-noise band approaches the noise floor set by amplifier, mixer, and readout electronics (\textcolor{matlabred}{\rule[0.75mm]{3mm}{.25mm}}).
We compute a thermal-noise-limited sensitivity (\textcolor{black}{\rule[0.75mm]{1.1mm}{.25mm}~\rule[0.75mm]{1.1mm}{.25mm}}) of 1.1~pT/$\sqrt{\text{Hz}}$ using the measured device response from Fig. \ref{fig:cavitysweeps}b.
See Supplemental Sec.~\ref{supp:noiseshape} for a discussion of the noise spectrum shape.
}
\label{fig:sensitivity}
\end{figure}

We measure a noise floor of $e_n=26$~$\mathrm{nV}/\sqrt{\mathrm{Hz}}$ at the digitizer under optimal operating conditions.
From this measurement, we project an optimal sensitivity of ${\eta=9.7\;\mathrm{pT}/\sqrt{\text{Hz}}}$ at 5~kHz  (Fig.~\ref{fig:sensitivity}).
In future work, DC signals could be upmodulated to this low-noise band using an AC bias field.

The measured sensitivity is limited by phase noise from the MW source, so sensitivity would improve with a lower phase noise MW source (see Supplemental Sec.~\ref{supp:thermalnoise}).
However, sensitivity improvements through modifications to the readout and MW electronics are fundamentally limited by the thermal-noise-limited sensitivity $\eta_{\mathrm{th}}= 1.1$~$\mathrm{pT}/\sqrt{\mathrm{Hz}}$ (see Supplemental Sec.~\ref{supp:thermalnoise}).
The sensitivity could also be enhanced by increasing the collective cooperativity parameter $\xi$, which can be improved by increasing the number of polarized spins, reducing the spin resonance linewidth, or by increasing the cavity quality-factor.


\section{Discussion}


This work demonstrates a generalization of the MW cavity readout technique combined with spin states prepared by thermal spin polarization arising from a zero-field splitting.
By removing the requirement of optical polarization, solid-state spin sensors can be made smaller, lighter and more power efficient.
Furthermore, the challenges associated with optical excitation are avoided, including light delivery; high heat loads; and laser pointing, polarization, and amplitude noise.

The technique demonstrated in this work is applicable to a broad class of paramagnetic defects, including those with a $\gtrsim$ GHz-scale ZFS.
As a result, paramagnetic defects which are currently unsuitable for sensors due to poor optical initialization or poor optical readout may become viable candidates for high-performance bulk solid-state sensing.
For example, the approach demonstrated here is expected to work well with silicon-vacancy or -divacancy in silicon carbide \cite{simin2015highprecision,kraus2014room,niethammer2016vector,abraham2021nanoTesla}.
Access to a broader range of defects may offer a varienty of advantages, including improved performance due to application-tailored energy level structures, material loss tangents, and dielectric constants.
In addition, the use of new defects could spur advances in sensing modalities beyond magnetometry, including electric field and time sensing \cite{trusheim2020clock}.

\clearpage
\bibliography{thebib.bib}

\clearpage

\onecolumngrid

\section{Supplement}
\subsection{Ruby hamiltonian eigenstates}
\label{supp:hamiltonian}
\begin{figure}[b]
\begin{minipage}[b]{0.45\textwidth}
\begin{overpic}[width=3.2in]{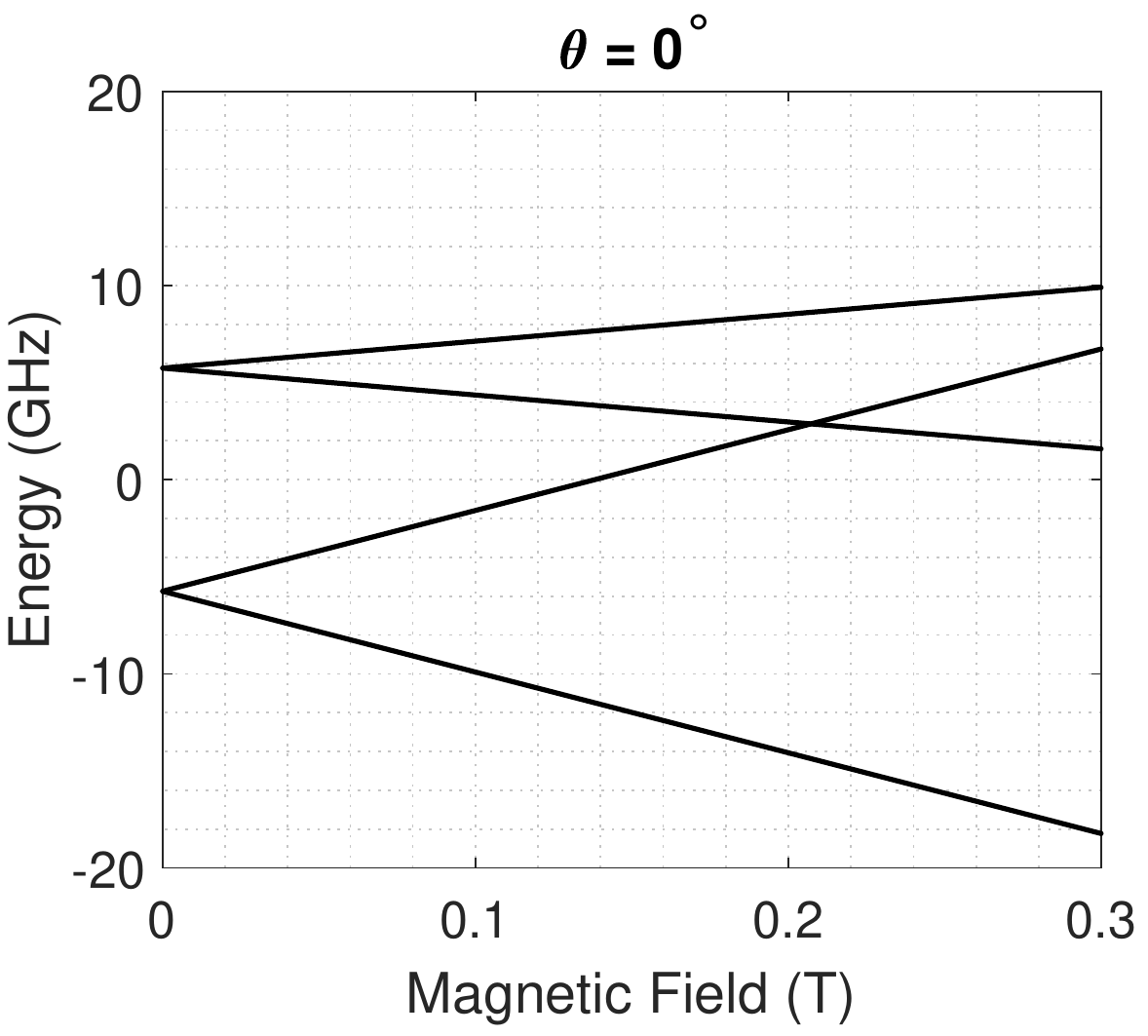} \put(0,85){\textbf{a)}}
\end{overpic}
\begin{overpic}[width=3.2in]{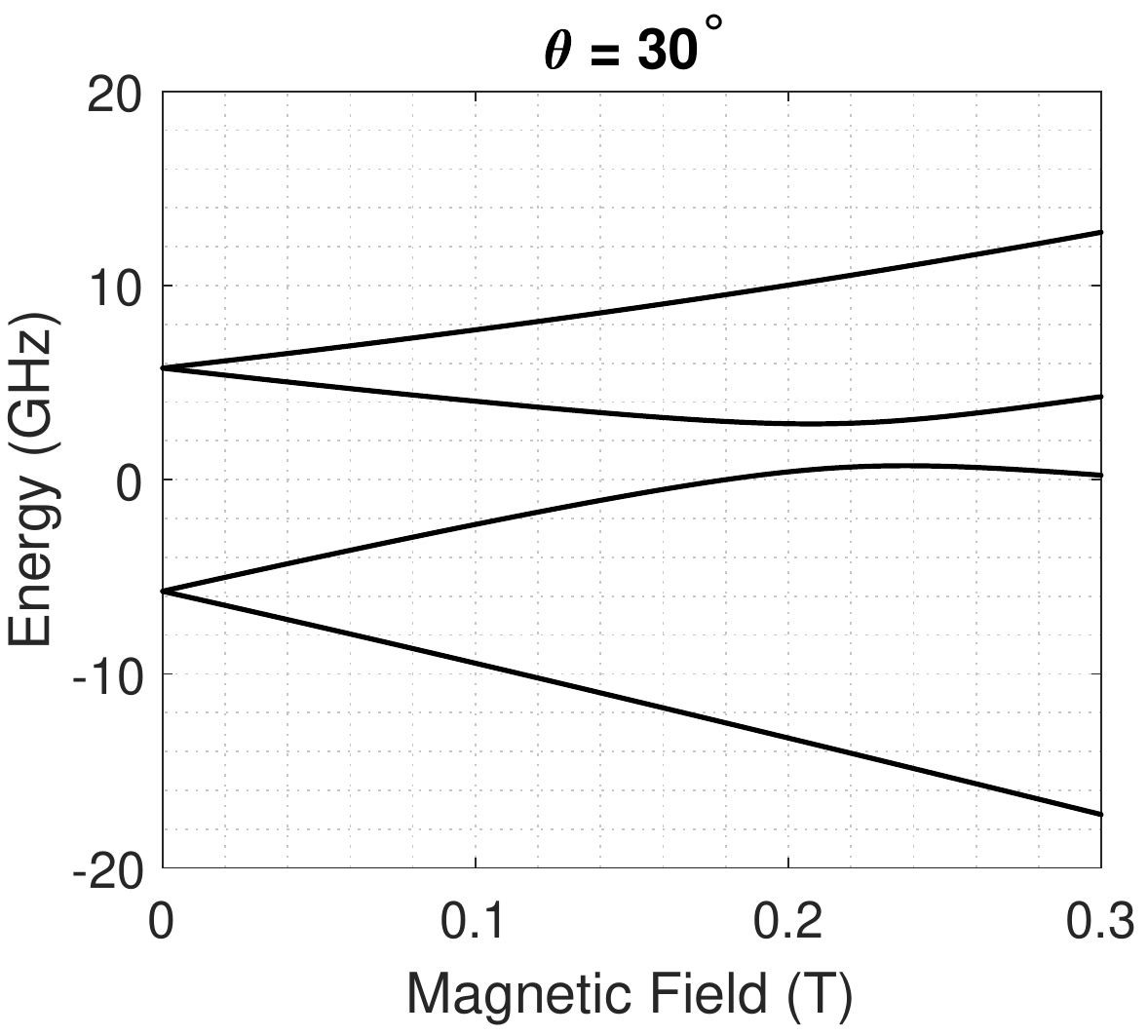} \put(0,85){\textbf{b)}}
\end{overpic}
\end{minipage}%
\;
\begin{minipage}[b]{0.45\textwidth}
\begin{overpic}[width=3.2in]{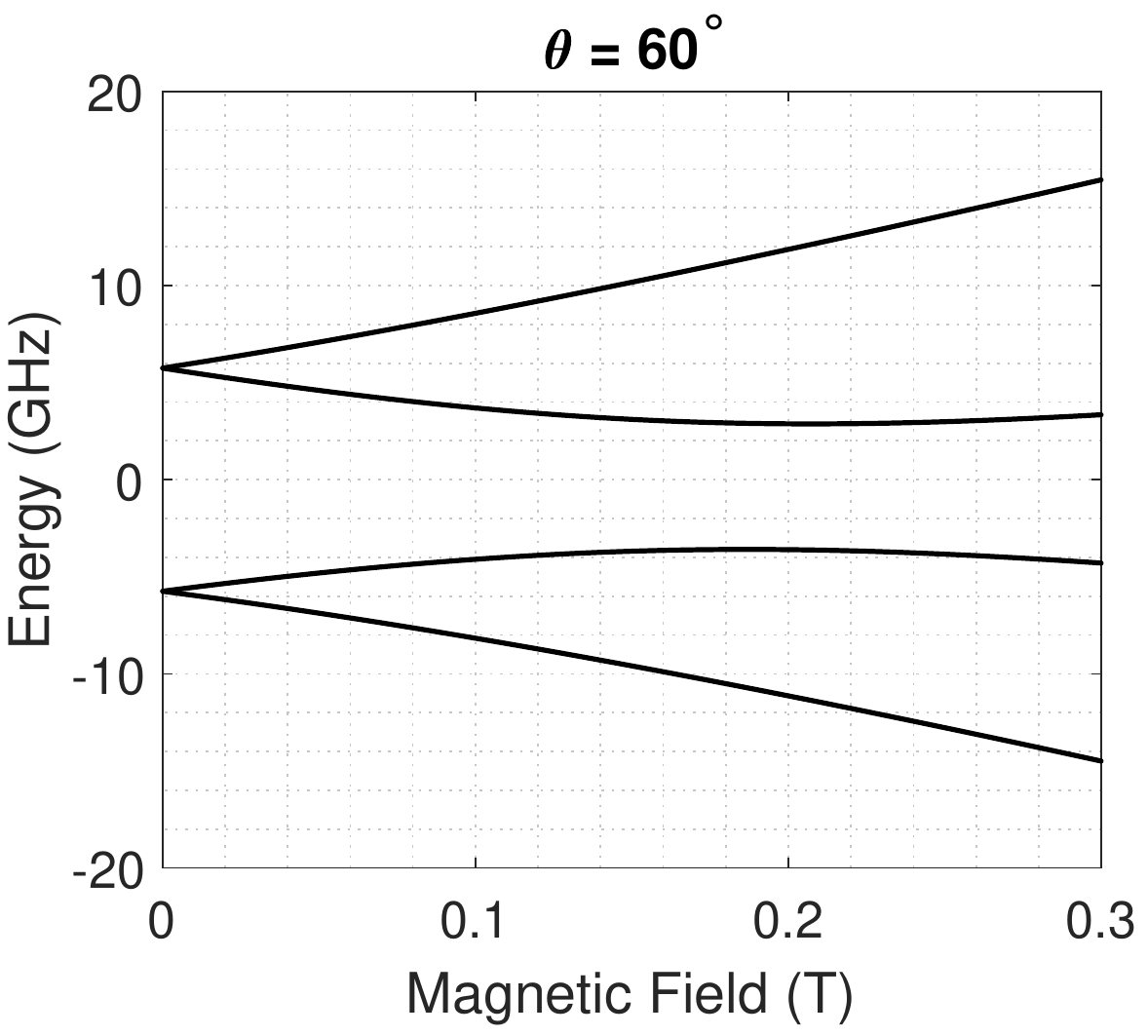} \put(0,85){\textbf{c)}}
\end{overpic}
\begin{overpic}[width=3.2in]{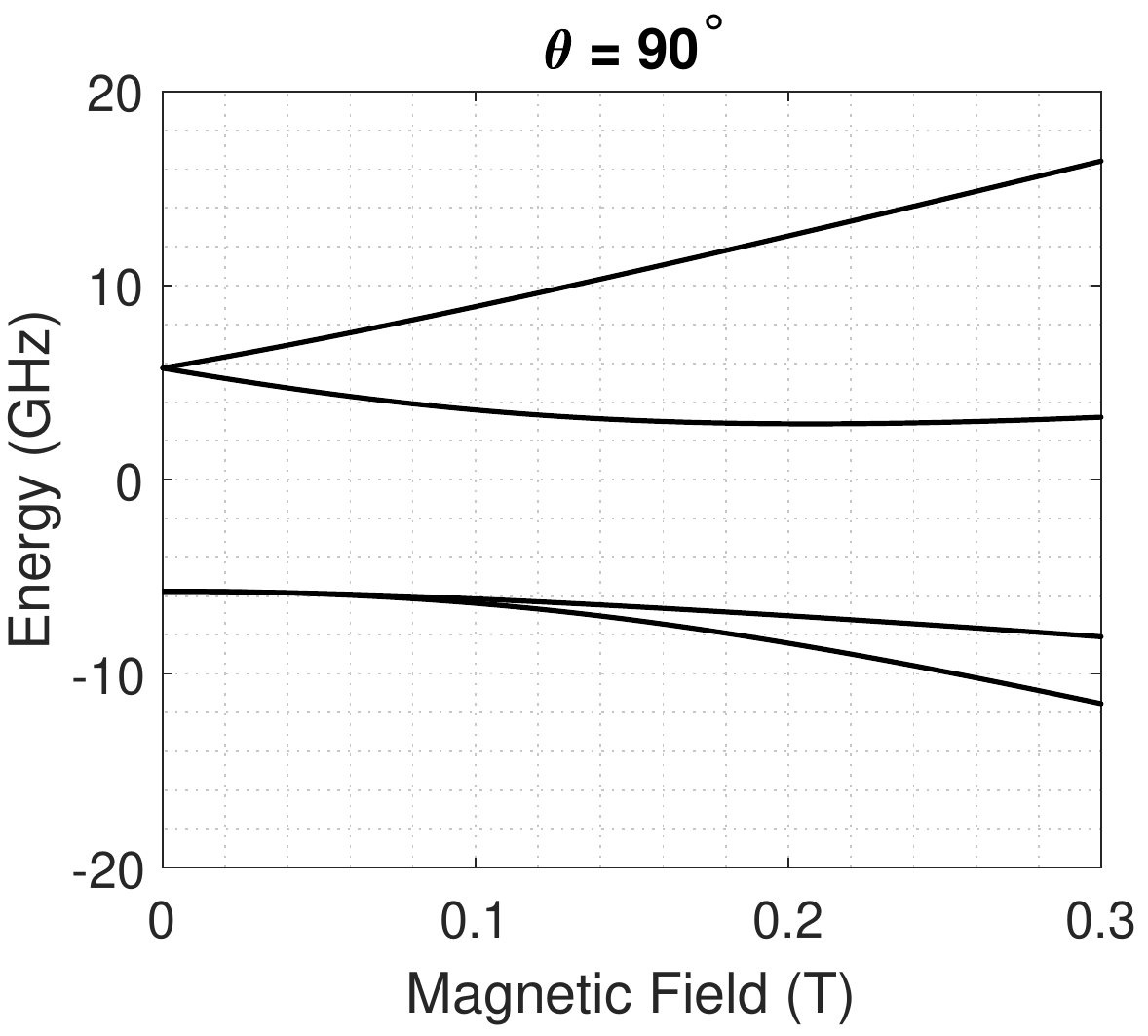} \put(0,85){\textbf{d)}}
\end{overpic}
\end{minipage}%
\caption{\textbf{Eigenenergies of the Cr$^{3+}$ for various magnetic field orientations.}}\label{fig:energylevels}
\end{figure}
Under application of a magnetic field $\mathbf{B} = B_z\hat{z}$ along the c-axis of the ruby, the eigenstates are $|m_s = \pm\tfrac{1}{2}\rangle$ and $|m_s = \pm\tfrac{3}{2}\rangle$, with eigenenergies \cite{rubynist1977}
\begin{equation}
\label{eqn:energies}
E_{\pm 1/2} = -D \pm \frac{1}{2}g_{||}\mu_B\hbar,\;\; E_{\pm 3/2} = D \pm \frac{3}{2}g_{||}\mu_B\hbar.
\end{equation}
As $D < 0$ \cite{rubynist1977}, the $|m_s = \pm\tfrac{1}{2}\rangle$ states are higher energy than the $|m_s = \pm\tfrac{3}{2}\rangle$ states at $B_z = 0$ (and for all $|B| < 2|D|/g_{||}\mu_B\hbar\approx 0.2$~T).
For magnetic field orientations not aligned along the c-axis, the pure $|m_s\rangle$ eigenstates mix together so that the eigenenergies no longer vary linearly in magnetic field.
The eigenenergies are calculated numerically for four distinct magnetic field orientations in Fig.~\ref{fig:energylevels}.
For more details, see Ref.~\cite{schulz-du1959}.

Two features are important to note about the resulting eigenstates show in Fig.~\ref{fig:energylevels}.
First, the allowed transition energy slope verus applied magnetic field is largest at $\theta = 0^{\circ}$ and $\theta = 90^{\circ}$; consequentially, magnetometry should be most sensitive at these orientations.
Numerical simulations corroborate this observation and predict the sensitivity at $\theta = 0^{\circ}$ and $\theta = 90^{\circ}$ orientations to be within 1\% of each other.
However, the $\theta = 90^{\circ}$ case requires precise alignment of the bias field ($< 1^{\circ}$) to remain comparable in sensitivity to the $\theta = 0^{\circ}$ case, which is difficult to ensure with the current experimental apparatus.
Second, because the eigenstates for $\theta \neq 0^{\circ}$ are mixtures of the pure $|m_s\rangle$ states, there exist allowed transitions that were forbidden in the $\theta = 0^{\circ}$ case.
These unwanted interactions near the desired transition can cause a highly non-linear dependence on applied magnetic field, including local extrema where the dependence on magnetic field is greatly reduced.
For these two reasons, we choose to align the magnetic field along the ruby c-axis when the device is operated as a magnetometer. 

\subsection{Modal field volume and filling factor calculation}
\label{supp:ff}

The modal field volume $V_\text{cav}$ is defined as
\begin{equation} \label{eq:Vcav}
V_\text{\text{cav}} = \frac{\int_{\text{all space}} |\mathbf B (\mathbf r)|^2 dV}{|\mathbf B_\text{max}|^2}.
\end{equation}
In Ansys HFSS, and through analytical calculations found in \cite{eisenach2021cavity}, we compute the modal field volume to be $V_\text{\text{cav}} = 52.2$~mm$^{3}$.
The physical volume of the ruby is $224$~mm$^3$, indicating that the applied MWs are well-contained within the ruby and justifying the use of the modal field volume to approximate the number of spins interacting with the MWs.
The filling factor (related to Eq.~\eqref{eq:Vcav} above) is defined as \cite{poole1996electron}
\begin{equation} \label{eq:FF}
\zeta = \frac{\int_{\text{ruby resonator}} |\mathbf B (\mathbf r)|^2 dV}{\int_{\text{all space}} |\mathbf B (\mathbf r)|^2 dV}.
\end{equation}
\noindent
Using HFSS we compute the filling factor of the ruby resonator to be $\zeta = 0.69$.
The HFSS model used to estimate $\zeta$ includes the SiC substrate used for thermal and mechanical stabilization.  

\subsection{Spin polarization calculations}
\label{supp:thermal}
In thermal equilibrium, the probability of occupying a state with energy $E_i$ is given by the Boltzmann distribution,
\begin{equation}
p_i = \frac{e^{-E_i/k_B T}}{\sum_j e^{-E_j/k_B T}},
\label{eqn:boltzmann}
\end{equation}
where $k_B$ is Boltzmann's constant and $T$ is the temperature, taken to be 293 K at room temperature.
We neglect the Zeeman energy shift, which at the bias field used (31 G), is negligible compared to the ZFS.
Taking $p_1$ and $p_2$ to refer to the $|\pm\tfrac{3}{2}\rangle$ states, we have $E_1 = E_2 = D \approx -5.75$~GHz, and with $p_3$ and $p_4$ referring to the $|\pm\tfrac{1}{2}\rangle$ states, $E_3 = E_4 = -D \approx 5.75$~GHz.
Equation~\eqref{eqn:boltzmann} then yields $p_1 = p_2 = 0.25024$ and $p_3 = p_4 = 0.24976$. We only address the $|+\tfrac{3}{2}\rangle \leftrightarrow |+\tfrac{1}{2}\rangle$ transition, and therefore only consider $p_1$ and $p_3$.
Furthermore, because we can only observe population differences between spin states, the effective polarization is $\mathcal{P} = |p_1 - p_3| = 4.7\times 10^{-4}$ of the total Cr$^{3+}$ population.
In this work the $|+\tfrac{3}{2}\rangle \leftrightarrow |+\tfrac{1}{2}\rangle$ and  $|-\tfrac{3}{2}\rangle \leftrightarrow|-\tfrac{1}{2}\rangle$ transitions are spectroscopically resolved, and probe MWs are chosen to address the positive $m_s$ states, so that negative $m_s$ states do not contribute to the effective polarization.
Since only half of the states are addressed, the effective polarization fraction is half as large as it would be if all of the states were used.

The total number of spins addressed by the interrogation MWs is
\begin{equation}
\label{eq:totalspins}
N_{\mathrm{tot}} = \frac{V_{\mathrm{cav}}}{V_{\mathrm{cell}}}\alpha_{\mathrm{Cr}^{3+}}\frac{m_{\mathrm{Al}_2\mathrm{O}_3}}{m_{\mathrm{Cr}_2\mathrm{O}_3}} N_{\mathrm{cell}},
\end{equation}
where $V_{\mathrm{cav}} = 52.2$~mm$^3$ is the modal field volume (see Supplemental Sec.~\ref{supp:ff}), $V_{\mathrm{cell}} = 0.2548$~nm$^3$ is the unit cell volume \cite{Villars2016:sm_isp_sd_1401045}, $\alpha_{\mathrm{Cr}^{3+}}=0.05$\% is the Cr$^{3+}$ concentration by weight (measured as the percent by weight of Al$_2$O$_3$ replaced with Cr$_2$O$_3$), $m_{\mathrm{Al}_2\mathrm{O}_3}$ and $m_{\mathrm{Cr}_2\mathrm{O}_3}$ are the molecular weights of Al$_2$O$_3$ and Cr$_2$O$_3$, and $N_{\mathrm{cell}} = 12$ is the number of Al atoms per unit cell \cite{Villars2016:sm_isp_sd_1401045}.
The calculation yields $N_{\mathrm{tot}} = 8\times 10^{17}$ Cr$^{3+}$ spins being interrogated.
The fit to the data in Fig.~\ref{fig:avoidedcrossing} suggests $N \approx 3.5\times 10^{14}$ polarized Cr$^{3+}$ spins between the $|m_s = +\tfrac{3}{2}\rangle$ and $|m_s = +\tfrac{1}{2}\rangle$ states. The resulting effective polarization is then $4.4\times 10^{-4}$, in agreement with the expected value from the Boltzmann distribution.

For comparison with optical polarization, we compute the laser power required to obtain the same number of polarized spins in a hypothetical scenario where ruby can be optically polarized and thermal polarization is neglected. The required optical power is estimated to be
\begin{equation}
\label{eq:opticalpower}
P = N\hbar\omega \kappa_{\mathrm{th}} n_{\mathrm{photons}}
\end{equation}
where $\omega$ is the frequency of light, $\kappa_{\mathrm{th}} = 1/T_1$ is the thermal polarization rate, $N$ is the number of polarized spins, and $n_{\mathrm{photons}}$ is the average number of photons required to polarize one spin.
We assume 532~nm green light, with $\omega = 2\pi\times 5.6\times 10^{14}$~Hz, and crudely guess $n_{\mathrm{photons}} = 3$ photons to polarize each spin.
For ruby, we fit a value of $\kappa_{\mathrm{th}} = 2\pi\times 120$~kHz (see Supplemental Sec.~\ref{supp:parameterestimation}). Therefore, we crudely estimate that optical polarization of the $N = 3.5\times 10^{14}$ spins would hypothetically require approximately $P = 300$~W.

\subsection{Magnetic field calibration}
\label{supp:testfieldamplitude}
In order to measure the device's sensitivity, a test magnetic field of known magnitude is applied.
The test field is generated by a low-noise signal generator driving a coil in series with a 50 $\Omega$ load, resulting in a total equivalent series resistance of 51.1 $\Omega$. The test field is generated by by a 1 Vpp signal at 10 Hz, resulting in an RMS current of $I = 6.9$~mA. The coil has a mean radius of $R=15.68$~mm and is made of $N_{\mathrm{turn}}=8$ turns of 1.35 mm diameter copper wire. The coil is placed with its center $z = 30$~mm from the ruby resonator center.

We determine the test field strength using three different methods: using a commercial magnetometer; using the known coil geometry and applied current; and using the response of the resonator to a calibrated bias magnetic field. The methods and their results are summarized in Table~\ref{tab:magfield}.

\begin{table}[t]  
\centering
\caption{Characterization of the test field amplitude} 
\centering 
\begin{tabular}{l c c}
\hline\hline   
Method & Amplitude (pT RMS) & Difference  \\
\hline
Commercial magnetometer & 242 & -    \\
Theoretical calculation & 220 & -9\%  \\
FEMM simulation & 233 & -4\% \\
Bias field sweep & 216 & -11\% \\
\hline 
\end{tabular}\label{tab:magfield}
\end{table}  

For the first method, we place a commercial magnetometer (TwinLeaf VMR) 30 mm from the test coil, which is the measured distance between the test coil and the ruby sample.
This measurement observes an RMS field of 242~nT.

For the second method, we use both an analytic calculation and a computer simulation to determine the applied magnetic field. The magnetic flux density along the axis of a solenoid is given approximately by
\begin{equation}
B = N_{\mathrm{turn}}\frac{\mu_0 I R^2}{2(z^2 + R^2)^{3/2}},
\label{eqn:magfieldtheory}
\end{equation}
where $\mu_0$ is the permeability of free space, $N_{\mathrm{turn}}$ is the number of turns, $I$ is the current through the coil, $R$ is the mean radius, and $z$ is the axial distance from the coil center. Using the coil geometry previously described, we compute 220 nT RMS. We also use the Finite Element Method Magnetics (FEMM) software package to simulate this same geometry and compute 233 nT RMS \cite{femm}. The difference in values is mostly because the the analytic calculation does not account for the finite spacing of the coils.

Finally, we compute the test field strength from the slope of the dispersive spin resonance signal by varying the bias magnetic field from the electromagnet itself.
To accurately measure the slope, the bias magnetic field strength must first be known.
To calibrate the bias field strength, the ruby resonator assembly is removed and a commercial magnetometer (Metrolab THM1176) is placed in the center of the electromagnet.
As the applied current is varied, the magnetic field is recorded using the commercial magnetometer.
The resulting plot of current versus magnetic field is linear.
Linear regression gives the desired mapping from applied current to magnetic field.

Once the bias magnetic field is calibrated, it is then swept across the cavity resonance, as in Fig.~\ref{fig:cavitysweeps}b.
From this data, we compute a maximal slope of the dispersive component of
$M_{\mathrm{max}}= 2994$~V/T.
We then perform a magnetometry measurement using the test field and look at the spectrum, similar to Fig.~\ref{fig:sensitivity}, to find the amplitude of the 10~Hz peak in the spectrum, which is $V_{m}$ = 0.646 mV RMS.
The estimated test field strength is then
\begin{equation}
B_{\mathrm{test}} = \frac{V_{m}}{M_{\mathrm{max}}},
\end{equation}
which evaluates to 216 nT RMS.

All four resulting values agree to within 11\% of the commercial magnetometer. For magnetometry measurements, we use the commercial magnetometer result of 242~nT.

\subsection{Reflection coefficient fitting}
\label{supp:parameterestimation}
The 2D data from Fig.~\ref{fig:avoidedcrossing}a was fit to Equations \eqref{eqn:reflection} and \eqref{eqn:spininteraction}.
From this data, we estimate the physical parameters $\kappa_c$, $\kappa_{c1}$, $\kappa_s$, $\kappa_{\mathrm{th}}$, and $g_{\mathrm{eff}} \equiv g_s\sqrt{N}$.
The number of MW photons in the cavity depends on $\kappa_c$, which must be introduced into the equation as $n_{\mathrm{cav}} = \tfrac{P}{\hbar\omega_d \kappa_c}$.
Note that $\hbar$, $\omega_d$ and  the applied MW power $P$ are known parameters and do not require fitting.
After collection, the 2D data are normalized to unity before being fit to the model of the complex reflection coefficient.

To account for slight experimental non-idealities, additional parameters are added to the fit.
The reflection coefficient amplitude is modified by a constant factor $A$, along with a component linear in $\omega_d$ with slope $b$ to account for cavity asymmetries.
To accommodate slight phase misalignment, a constant phase offset $\psi$ is added.
Additive offsets are accounted for with parameters $o_r$ and $o_i$ for the real and imaginary parts, respectively.
The MW frequency and spin resonance frequency have offsets $\omega_{s,\mathrm{off}}$ and $\omega_{d,\mathrm{off}}$ applied, respectively.
Due to a difference in electrical length of the signal and reference arm, a delay term $\tau$ is applied to model the phase rotation as a function of frequency.
A summary of the additional parameters being fit is given in Table~\ref{tab:params}, along with numerical values from the fit.
The final functional form being fit is given by
\begin{equation}
\label{eq:fiteq}
\Gamma'(\omega_s,\omega_d) = o_r + io_i + e^{i(\psi + (\omega_d - \overline{\omega_d}) \tau)}(1 + A + b(\omega_d- \overline{\omega_d}))\Gamma(\omega_s - \omega_{s,\mathrm{off}}, \omega_d - \omega_{d,\mathrm{off}}),
\end{equation}
where $\overline{\omega_d}$ is the mean value of $\omega_d$ over the data taken.

For fitting, both the real and imaginary components are used, with the objective function given by
\begin{equation}
\sum_{\omega_s,\omega_d}|\mathrm{Re}\{\Gamma'(\omega_s,\omega_d)\}| + |\mathrm{Im}\{\Gamma'(\omega_s,\omega_d)\}|.
\end{equation}
L1 norms were used instead of L2 norms because the former were found to yield more accurate data reconstruction.

\begin{table}[ht]  
\centering
\caption{Additional fit parameters to account for slight experimental non-idealities} 
\centering 
\begin{tabular}{l | l | c | l}
\hline\hline   
Name & Description & Fit value & Unit\\
\hline
$o_r$ & Offset in real reflection coefficient & -0.008 & unitless \\
$o_q$ & Offset in imaginary reflection coefficient & 0.12 & unitless    \\
$A$ & Amplitude correction of reflection coefficient & 0.003 & unitless    \\
$b$ & Cavity asymmetry slope & $10^{-9}$ & s  \\
$\psi$ & Phase offset & 0.14 & rad \\
$\tau$ & Delay & $-1.2\times 10^{-8}$ & s    \\
$\omega_{s,\mathrm{off}}$ & Offset in spin resonance frequency & $-7.3\times 10^{6}$ & rad/s \\
$\omega_{d,\mathrm{off}}$ & Offset in MW drive frequency & $-5.6\times 10^{5}$ & rad/s  \\
\hline 
\end{tabular}\label{tab:params}
\end{table}  

To ensure that the auxiliary parameters don't change the model appreciably, the parameters $o_r$, $o_i$, $A$, and $\psi$ should be small compared to 1.
Comparison of the numeric values found verifies that this is the case.
With the given range of $\omega_d$ (a span of $\sim$5~MHz), the fit value of $b$ results in an amplitude correction of $\sim$0.05.
The delay term $\tau$ is a physical parameter and should be consistent with the experimental setup.
From the value estimated for $\tau$, an electrical length of $\sim$3~m is calculated, which is roughly the length of cabling used in the experiment.

The values for $\kappa_c$, $\kappa_{c1}$, $\kappa_s$, and $g_{\mathrm{eff}}$ extracted from the data are found to be insensitive to the method of the fit and applied MW power.
However, the fit for $\kappa_{\mathrm{th}}$ is sensitive to the fitting method and applied MW power.
Specifically, fitting becomes very poor at lower applied MW powers (compared to $T_1$ values in \cite{T1corundum1960}).
This behavior can be understood by looking at the reflection coefficient $\Gamma$, which depends on $\kappa_{\mathrm{th}}$ through the interaction term $\Pi$ given in Eq.~\eqref{eqn:spininteraction}.
Because $\kappa_{\mathrm{th}}$ appears only once as the ratio $g_s^2n_{\mathrm{cav}}\kappa_s/(2\kappa_{\mathrm{th}})$, the accuracy of fitting $\kappa_{\mathrm{th}}$ depends on the accuracy of fitting both $\kappa_s$ and $\kappa_c$ (through $n_{\mathrm{cav}})$.
This suggests that the fit for $\kappa_{\mathrm{th}}$ would be noisier than that for the other parameters, which occur in other places in the reflection coefficient.

\begin{table}[ht]
\centering
\caption{Comparison of estimated relaxation times for Cr$^{3+}$ in Al$_2$O$_3$ at room temperature} 
\centering 
\begin{tabular}{l | c | c}
\hline\hline   
 & Current work & Manenkov et al (1960) \cite{T1corundum1960}  \\
\hline
$T_1$ & 1.3~$\mu$s & 2.4~$\mu$s \\
$T_2$ & 7.6~ns & 5.5~ns \\
\hline 
\end{tabular}\label{tab:times}
\end{table}  

The power dependence is also explained by this interaction term.
To see this, assume a small spin detuning so that $\kappa_s \gg |\omega_s - \omega_d|$.
The term dependent on $\kappa_{\mathrm{th}}$ is suppressed by $|\omega_s - \omega_d|$, so it is possible to observe the effect of $\kappa_{\mathrm{th}}$ only in the small detuning region.
Using the approximation $(1 - x)^{-1}\approx 1 + x$ for $|x| \ll 1$, the interaction term becomes
\begin{equation}
\label{eq:pi_approx}
\Pi \approx \frac{1}{\frac{\kappa_s}{2} + \frac{g_s^2n_{\mathrm{cav}}}{\kappa_{\mathrm{th}}}}\times\frac{g_s^2 N}{1 + \frac{2 i (\omega_d-\omega_s)}{\kappa_s}}.
\end{equation}
The dependence on $\kappa_{\mathrm{th}} = 1/T_1$ is entirely  captured in the first multiplicand.
To obtain an accurate estimate on $\kappa_{\mathrm{th}}$, the term $\tfrac{g_s^2n_{\mathrm{cav}}}{\kappa_{\mathrm{th}}}$ must therefore be on the same order or greater than $\kappa_s/2 = 1/T_2$.
Using the literature values $T_1 = 2.6$~$\mu$s and $T_2 = 5.5$~ns \cite{T1corundum1960} and solving  $\kappa_s/2 = \tfrac{g_s^2n_{\mathrm{cav}}}{\kappa_{\mathrm{th}}}$ results in an applied MW power of 2~dBm.
Thus, the fit found by applying 0~dBm of power should be able to estimate $\kappa_{\mathrm{th}}$, but any lower powers would become increasingly unreliable.
This agrees with the observation that fitting for applied MW powers of -5~dBm and lower results in fitting difficulties.
The fit was carried out at 0~dBm, which is the lowest power at which an accurate fit is possible.
The resulting $T_1$ and $T_2$ parameters are compared against results from \cite{T1corundum1960} and shown in Table~\ref{tab:times}, demonstrating reasonable agreement.

\subsection{Microwave power and bias field optimization}
\label{supp:power}
\begin{figure}[t]
\begin{minipage}[b]{0.45\textwidth}
\begin{overpic}[width=3.0in]{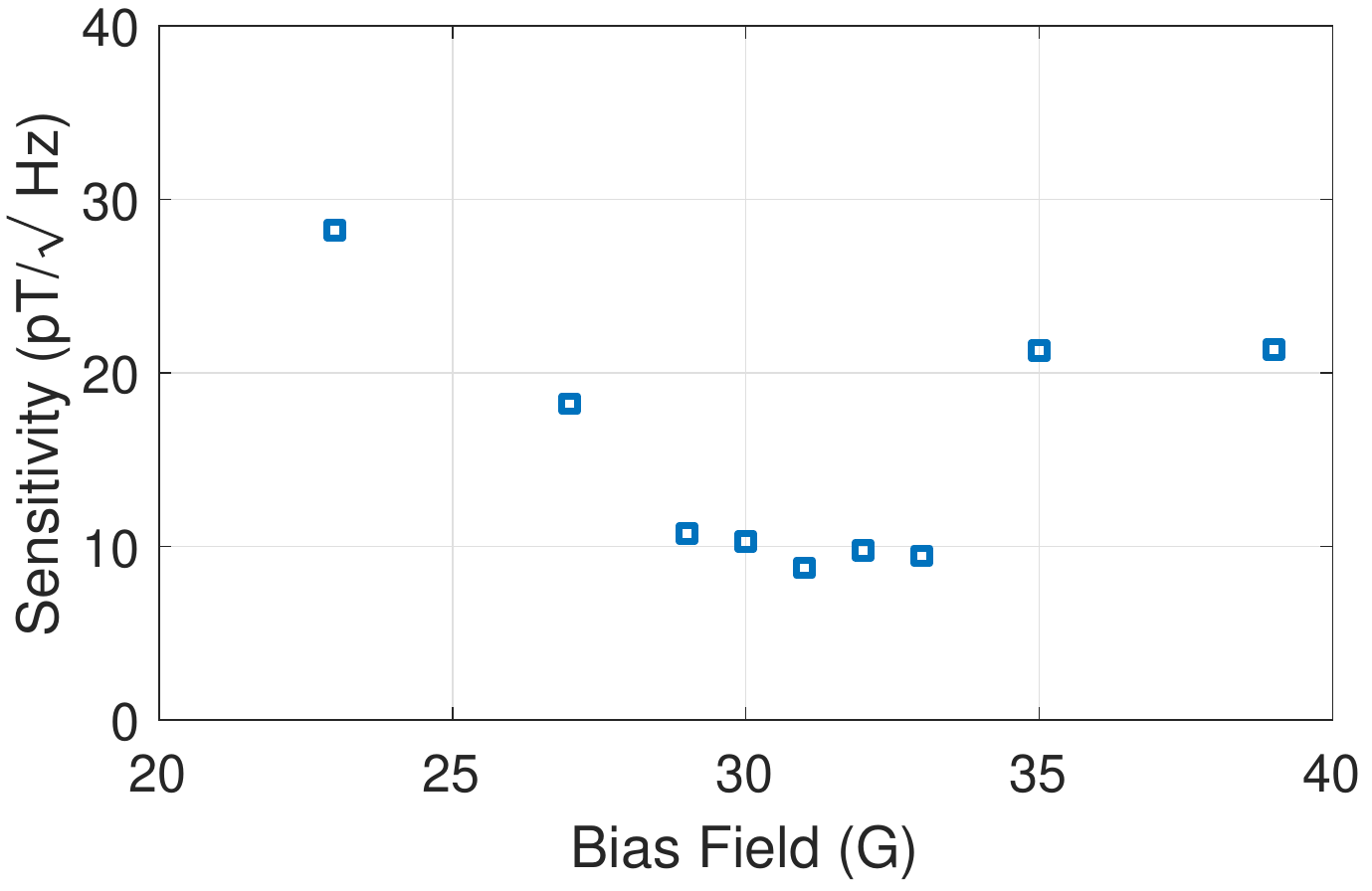} \put(0,65){\textbf{a)}}
\end{overpic}
\end{minipage}%
\;
\begin{minipage}[b]{0.4\textwidth}
\begin{overpic}[width=3.0in]{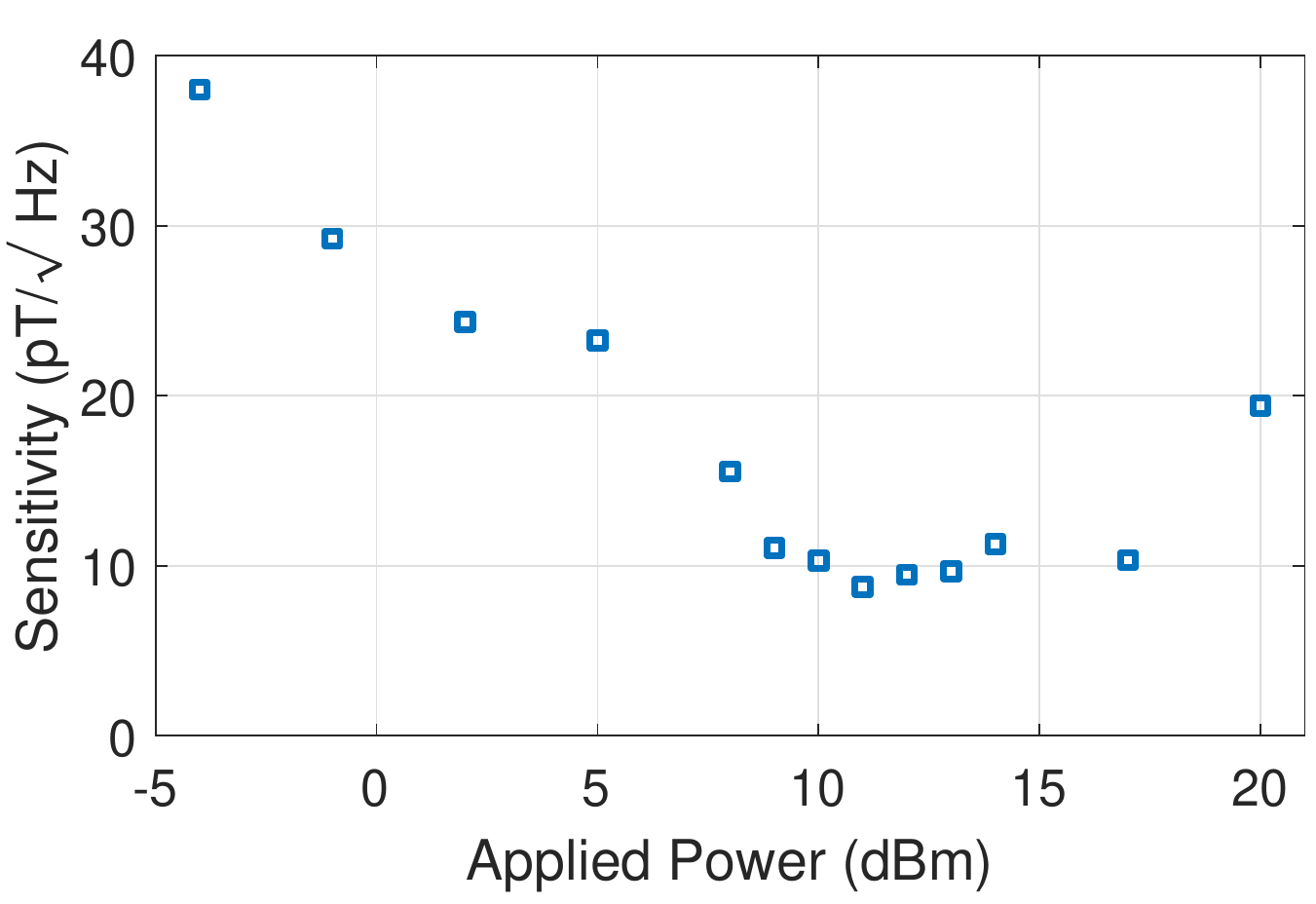} \put(0,65){\textbf{b)}}
\end{overpic}
\end{minipage}
\caption{\textbf{Bias magnetic field and power optimization.} a) The sensitivity is measured versus the bias magnetic field. We observe best sensitivity near 31~G. b) The sensitivity is measured versus interrogation MW power. Best sensitivity is observed with 11~dBm of interrogation power.}\label{fig:poweroptimization}
\end{figure}

\begin{figure}[t]
\begin{minipage}[b]{0.435\textwidth}
\begin{overpic}[width=3.0in]{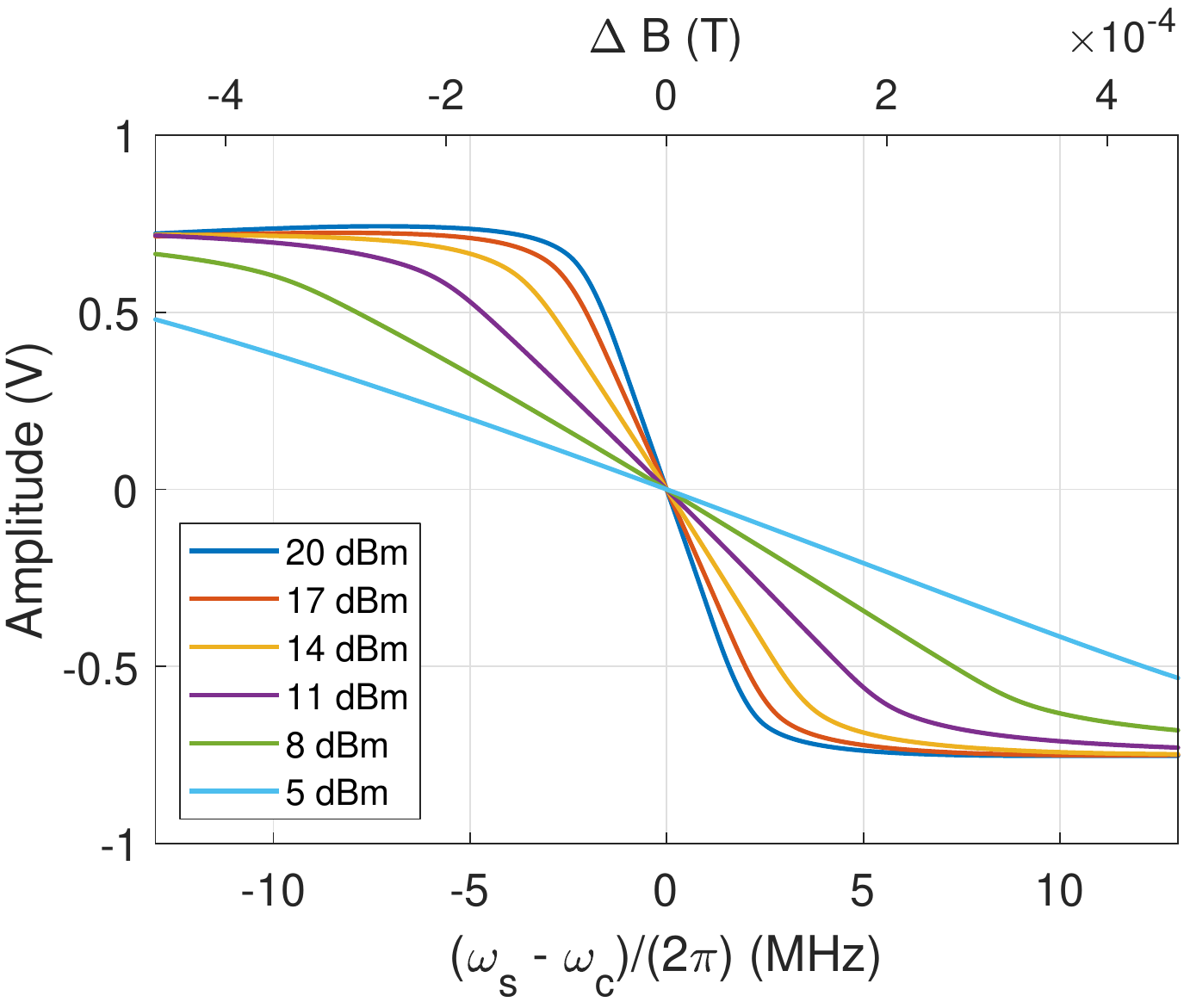} \put(0,80){\textbf{a)}}
\end{overpic}
\end{minipage}%
\;
\begin{minipage}[b]{0.435\textwidth}
\begin{overpic}[width=3.0in]{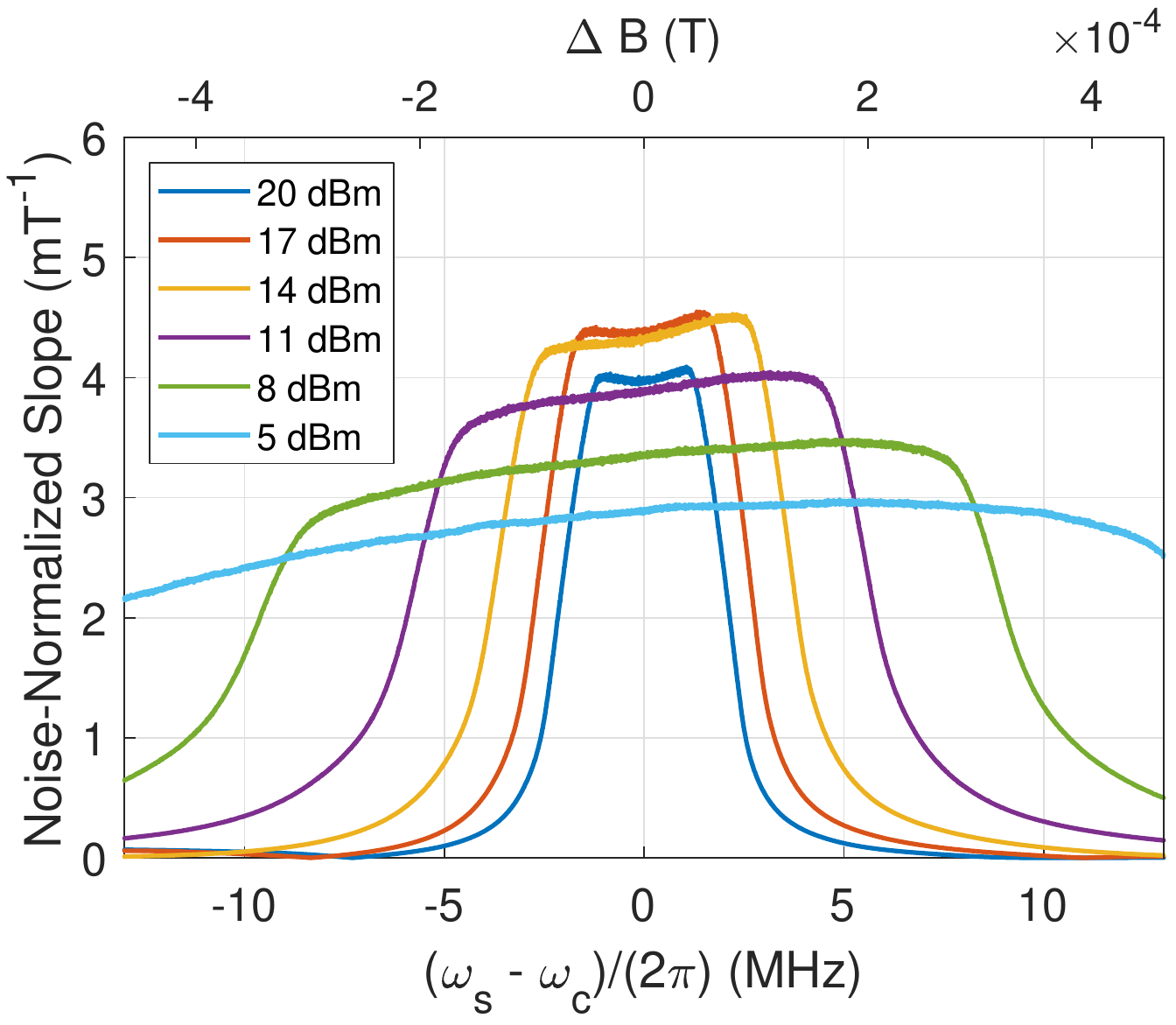} \put(0,80){\textbf{b)}}
\end{overpic}
\end{minipage}
\caption{\textbf{MW power variation.} a) The dispersive component of the reflected signal is plotted against the spin cavity detuning for a variety of applied powers to probe effect of applied power. The slope of the signal continually increases with power. b) The noise-normalized slope of the reflection coefficient, which is a good metric for overall sensitivity, is plotted against spin detuning for a variety of applied powers. The noise-normalized slope has an optimal power between 14 and 17~dBm, demonstrating increasing power cannot improve sensitivity indefinitely. The observed asymmetry about the resonance is attributed to interaction with the $|-\tfrac{3}{2}\rangle\leftrightarrow|-\tfrac{1}{2}\rangle$ resonance.}\label{fig:powervariation}
\end{figure}
For sharpest slope, and consequently highest sensitivity, the magnetic bias field should be configured so that the spin resonance frequency $\omega_s$ and the cavity resonance frequency $\omega_c$ are equal \cite{eisenach2021cavity}.
However, the optimal applied MW power is not immediately obvious: increasing the applied MW power increases the absolute reflected signal but also causes deleterious power broadening, which lowers the signal slope.

To optimize the bias magnetic field and applied MW power, we sweep both the bias magnetic field and the applied MW power and compute the sensitivity for each configuration.
For each bias magnetic field $B$ and applied MW power $P$, this gives a sensitivity $\eta(B, P)$.
The optimal sensitivity at a fixed bias magnetic field is then given by $\eta_{\mathrm{mag}}(B) = \min_P \eta(B, P)$.
The resulting plot of $\eta_{\mathrm{mag}}(B)$ as a function of bias field is shown in Fig.~\ref{fig:poweroptimization}a.
The bias field, which is calibrated using a commercial magnetometer (see Supplemental Sec.~\ref{supp:testfieldamplitude}), is swept from 23~G to 39~G and the MW power from -4~dBm to 20~dBm.
Using sensitivity as our figure of merit, we find the optimal bias field is 31 G, which agrees with $\omega_s = \omega_c$ based on ZFS data from \cite{rubynist1977}.
Furthermore, if the bias field is between 29~G and 33~G, the sensitivity remains nearly constant.
Note that sensitivity measurements were observed to fluctuate $\sim$2~pT/$\sqrt{\mathrm{Hz}}$ across repeated measurements due to ambient noise conditions and thermal resonance shifts moving the ruby resonance away from the drive frequency.
Thermal effects become more pronounced at higher powers.

To optimize the applied MW power, we instead compute $\eta_{\mathrm{MW}}(P) = \min_B\eta(B, P)$, which finds the optimal sensitivity for each applied MW power.
A plot of $\eta_{\mathrm{MW}}(P)$ as a function of applied MW power is shown in Fig.~\ref{fig:poweroptimization}b.
We find the optimal MW power applied to the cavity to be 11~dBm, with nearly flat sensitivity for MW powers between 9~dBm and 17~dBm.

We can further probe the effect of varying MW power by looking at the reflected dispersive signal as the bias magnetic field is swept across the cavity resonance (shown in Fig.~\ref{fig:powervariation}a).
The magnitude of the reflected signal increases as the power is increased, up until the low-noise amplifier is driven at its compression point.
The slope of this plot determines the amplitude of the 10 Hz peak in the spectrum observed in the sensitivity measurements.
The continually sharper slopes as applied MW power increases indicates that the observed spectral peak amplitude in sensitivity measurements increases with applied MW power.

We introduce a new metric, which we call the noise-normalized slope, defined by $M_{\mathrm{norm}} = M/\sqrt{PR}$, where $M$ is the slope of the dispersive reflected voltage as a function of applied magnetic field, $P$ is the applied MW power, and $R = 50$~$\Omega$ is the termination impedance.
Since the system is phase-noise-limited, we can approximate $e_n\propto \sqrt{PR}$.
The signal-to-noise ratio can then be approximated by $\mathrm{SNR}\propto M_{\mathrm{norm}}$.
By Eq.~\eqref{eqn:sensitivity}, $\eta\propto 1/\mathrm{SNR}$, so maximizing $M_{\mathrm{norm}}$ is roughly equivalent to minimizing sensitivity.
The noise-normalized slope is plotted for various applied MW powers in  Fig.~\ref{fig:powervariation}b.
Unlike the raw slope, there is an optimal applied MW power to maximize the noise-normalized slope, which is attributed to power broadening effects at higher power.
The data suggest the optimal MW power lies between 14 dBm and 17 dBm, which is roughly consistent with the results shown in Fig.~\ref{fig:poweroptimization}b.

\subsection{IQ Mixing}
\label{supp:iqmixing}
The signal generator output is $s(t) = e^{i\omega_d t}$.
After reflecting off the ruby cavity, the signal becomes $r(t) = \Gamma(\omega_d)e^{i\omega_d t + \phi_0}$, where $\phi_0$ is the phase delay of the reflected signal due to differing path lengths between the signal and reference arms.
Mixing the signals gives
\begin{equation}
\label{eq:mixer}
m(t) = r(t)s^*(t) = \Gamma(\omega_d)e^{i\phi_0}.
\end{equation}
The reflection coefficient $\Gamma$ contains the absorptive signal in its real component and the dispersive signal in its imaginary component.
The mixed signal $m(t)$ no longer has this property because of the phase $e^{i\phi_0}$.

As both the I and Q channels of the mixer are digitized, the full complex signal $m(t)$ is available for signal processing.
The complex multiplication by $e^{-i\phi}$ can therefore be performed in software, giving the final signal
\begin{equation}
\label{eq:phaserotation}
w(t) = e^{-i\phi t}m(t) = \Gamma(\omega_d)e^{i(\phi_0 - \phi)}
\end{equation}
Setting $\phi = \phi_0$ restores the reflection coefficient: $w(t) = \Gamma(\omega_d)$.
To distinguish this signal from the physical outputs of the IQ mixer, the real part of $w(t)$ is referred to as the absorptive channel and the imaginary part as the dispersive channel.

\subsection{Noise spectrum shape}
\label{supp:noiseshape}
\begin{figure}[t]
\begin{minipage}[b]{0.435\textwidth}
\begin{overpic}[width=3.0in]{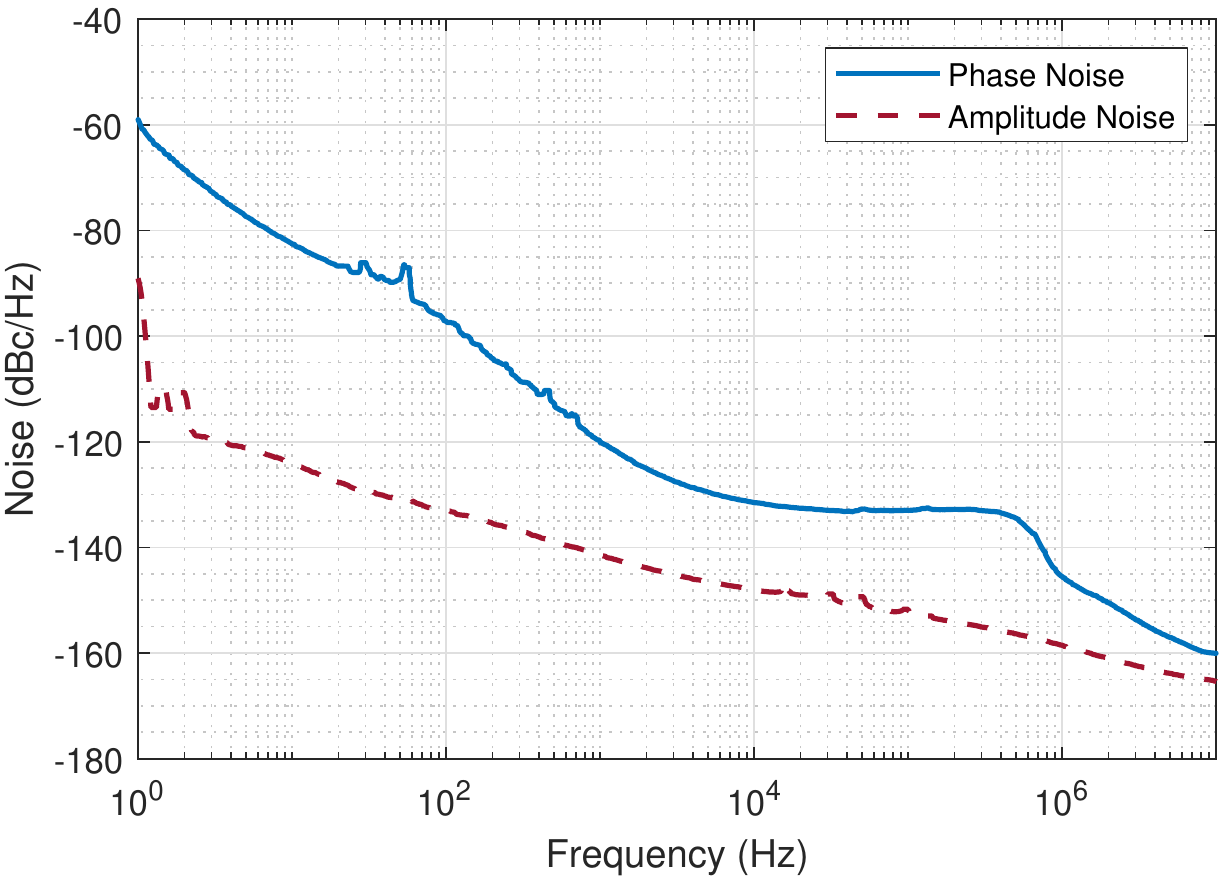} \put(0,70){\textbf{a)}}
\end{overpic}
\end{minipage}%
\;
\begin{minipage}[b]{0.435\textwidth}
\begin{overpic}[width=3.0in]{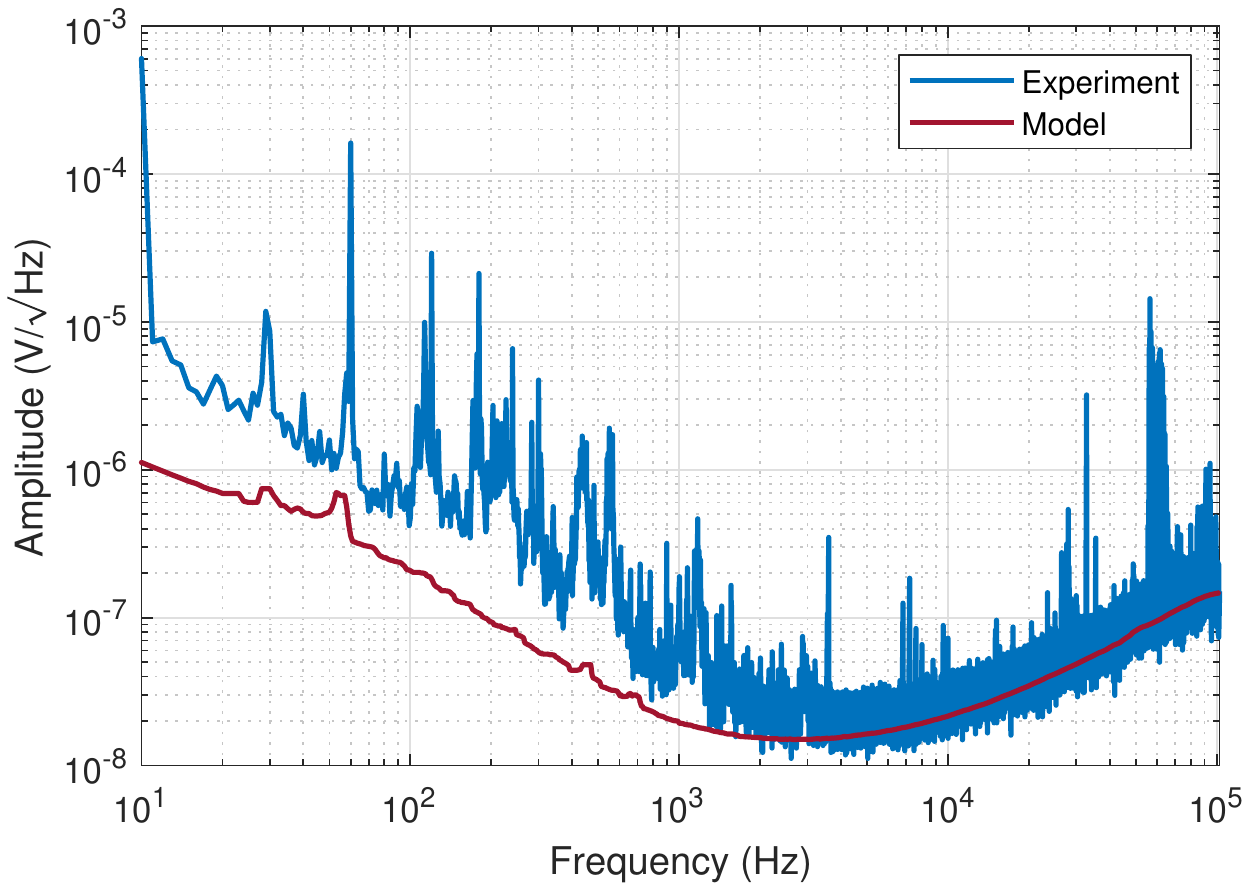} \put(0,70){\textbf{b)}}
\end{overpic}
\end{minipage}
\caption{\textbf{Noise spectra.} a) Measured phase noise (\textcolor{matlabblue}{\rule[0.75mm]{3mm}{.25mm}}) and amplitude noise (\textcolor{matlabred}{\rule[0.75mm]{1.1mm}{.25mm}~\rule[0.75mm]{1.1mm}{.25mm}}) of the probe MW source. b) Measured noise spectrum of the MW signal reflected off the ruby resonator (\textcolor{matlabblue}{\rule[0.75mm]{3mm}{.25mm}}) compared against the predicted noise spectrum (\textcolor{matlabred}{\rule[0.75mm]{3mm}{.25mm}}) given by Eq.~\eqref{eq:phasenoisepoweroutput}.}\label{fig:noiseshape}
\end{figure}
The shape of the reflected noise spectrum in Fig.~\ref{fig:sensitivity} can be explained by looking at the phase noise of the signal generator in combination with the frequency response of the reflection coefficient $\Gamma$.
Intuitively, the noise spectrum initially decreases with frequency because the signal generator's phase noise decreases with increasing offset frequency from the carrier.
This effect is eventually outweighed by the increase in the reflection coefficient as the probe MW frequency moves away from resonance.

Because we encode our signal on the imaginary dispersive channel, we are sensitive only to phase noise on the carrier.
To model the resulting noise on the output signal, we therefore need to understand how phase and amplitude noise transform through the system.
For simplicity, we assume our signals are centered at zero frequency.
That is, we mix down by the MW frequency $\omega_d$ so that $\Gamma(0)$ corresponds to the  reflection coefficient at $\omega_d$. Since amplitude and phase noise manifest as modulation on the carrier $\omega_d$, this allows us to treat these noise sources as baseband signals.

We decompose the reflection coefficient into real and imaginary parts, so $\Gamma(\omega) = \Gamma_{\mathrm{re}}(\omega) + i\Gamma_{\mathrm{im}}(\omega)$ where $\Gamma_{\mathrm{re}}$ and $\Gamma_{\mathrm{im}}$ are both real. Define
\begin{align}
\label{eq:gammasymmetry}
\Gamma_p(\omega) = \frac{\Gamma_{\mathrm{re}}(\omega) + \Gamma_{\mathrm{re}}(-\omega)}{2} + \frac{i\Gamma_{\mathrm{im}}(\omega) - i\Gamma_{\mathrm{im}}(-\omega)}{2} \\
\label{eq:gammaasymmetry}
\Gamma_s(\omega) = \frac{\Gamma_{\mathrm{re}}(\omega) - \Gamma_{\mathrm{re}}(-\omega)}{2} + \frac{i\Gamma_{\mathrm{im}}(\omega) + i\Gamma_{\mathrm{im}}(-\omega)}{2},
\end{align}
so that $\Gamma(\omega) = \Gamma_p(\omega) + \Gamma_s(\omega)$. The inverse Fourier transform of a real, symmetric signal is real, while that of a real, anti-symmetric signal is imaginary. Equations \eqref{eq:gammasymmetry} and \eqref{eq:gammaasymmetry} thus have the property that the inverse Fourier transform $\gamma_p(t)$ is real and $\gamma_s(t)$ is imaginary.
Therefore, $\Gamma_p(\omega)$ preserves phase noise as phase noise and amplitude noise as amplitude noise, while $\Gamma_s(\omega)$ swaps amplitude noise to phase noise and vice-versa.

The time-domain MW probe signal is written as $s(t) = 1 + \alpha(t) + \phi(t)$, where $\alpha(t)$ is the amplitude noise and $\phi(t)$ is the phase noise, assuming $|\alpha(t)|\ll 1$ and $|\phi(t)|\ll 1$. In the frequency-domain, this is $S(\omega) = 2\pi\delta(\omega) + A(\omega) + i\Phi(\omega)$. The resulting signal after reflecting off the resonator and going through the mixer is
\begin{equation}
r(t) = [(1 + \alpha(t) + i\phi(t))\ast\gamma(t)](1 - i\phi(t))
\end{equation}
where $1 - i\phi(t)$ corresponds to noise from the mixer and $\ast$ is the convolution operator.
Only phase noise propagates through the mixer since it is driven in saturation and amplitude noise is suppressed.
The corresponding frequency-domain signal is then
\begin{equation}
R(\omega) = 2\pi\delta(\omega)\Gamma(\omega) + A(\omega)\Gamma(\omega) + i\Phi(\omega)\Gamma(\omega) - i\Phi(\omega)\Gamma(0) - \frac{1}{2\pi} i [A(\omega)\Gamma(\omega)]\ast \Phi(\omega) + \frac{1}{2\pi}[\Phi(\omega)\Gamma(\omega)]\ast \Phi(\omega)
\end{equation}

We want an expression for the noise, so we remove $2\pi\delta(\omega)\Gamma(\omega)$ which corresponds to the noise-free signal.
Furthermore, we can ignore higher order products of noise since we assume the noise to be small relative to the signal.
This gives the expression for the output noise signal
\begin{equation}
N(\omega) = A(\omega)\Gamma(\omega) + i\Phi(\omega)\Gamma(\omega) - i\Phi(\omega)\Gamma(0)
\end{equation}
We need only consider the phase noise on the output signal, which is given by
\begin{equation}
\label{eq:phasenoiseoutput}
N_{\phi}(\omega) = A(\omega)\Gamma_s(\omega) + i\Phi(\omega)\Gamma_p(\omega) - i\Phi(\omega)\Gamma_p(0)
\end{equation}

All of the terms in the above equation are independent. The first is amplitude noise and is independent of the other two because they depend on phase noise. The last two are independent because $i\Phi(\omega)\Gamma_p(\omega)$ corresponds to phase noise reflected off the cavity, while $i\Phi(\omega)\Gamma_p(0)$ corresponds to the noise-free carrier reflected off the cavity with phase noise added on by the mixer, which has enough of a time delay relative to the reflected signal to be considered independent. To take the power spectrum, we can therefore add each term in quadrature. The final power spectrum is then
\begin{equation}
\label{eq:phasenoisepoweroutput}
P_{\phi}(\omega) = |A(\omega)\Gamma_s(\omega)|^2 + |\Phi(\omega)\Gamma_p(\omega)|^2 + |\Phi(\omega)\Gamma_p(0)|^2 + P_0
\end{equation}
where $P_0$ accounts for digitizer, thermal, and amplifier noise, which has a flat spectrum.

Using Eq.~\eqref{eq:phasenoisepoweroutput}, we can then predict the shape of the noise spectrum based on the frequency response of the cavity, seen in Fig.~\ref{fig:cavitysweeps}a, and noise spectrum of the Rhode \& Schwarz SMA100B-B711 signal generator, seen in Fig.~\ref{fig:noiseshape}a. The resulting prediction is compared against the experimental results in Fig.~\ref{fig:noiseshape}b, demonstrating a good prediction of noise at high frequency. Disagreement at low noise could be due to external noise sources, such as acoustic noise, or imperfect isolation in the circulator and mixer.

\begin{figure}[t]
\centering
\includegraphics[width=3.375in]{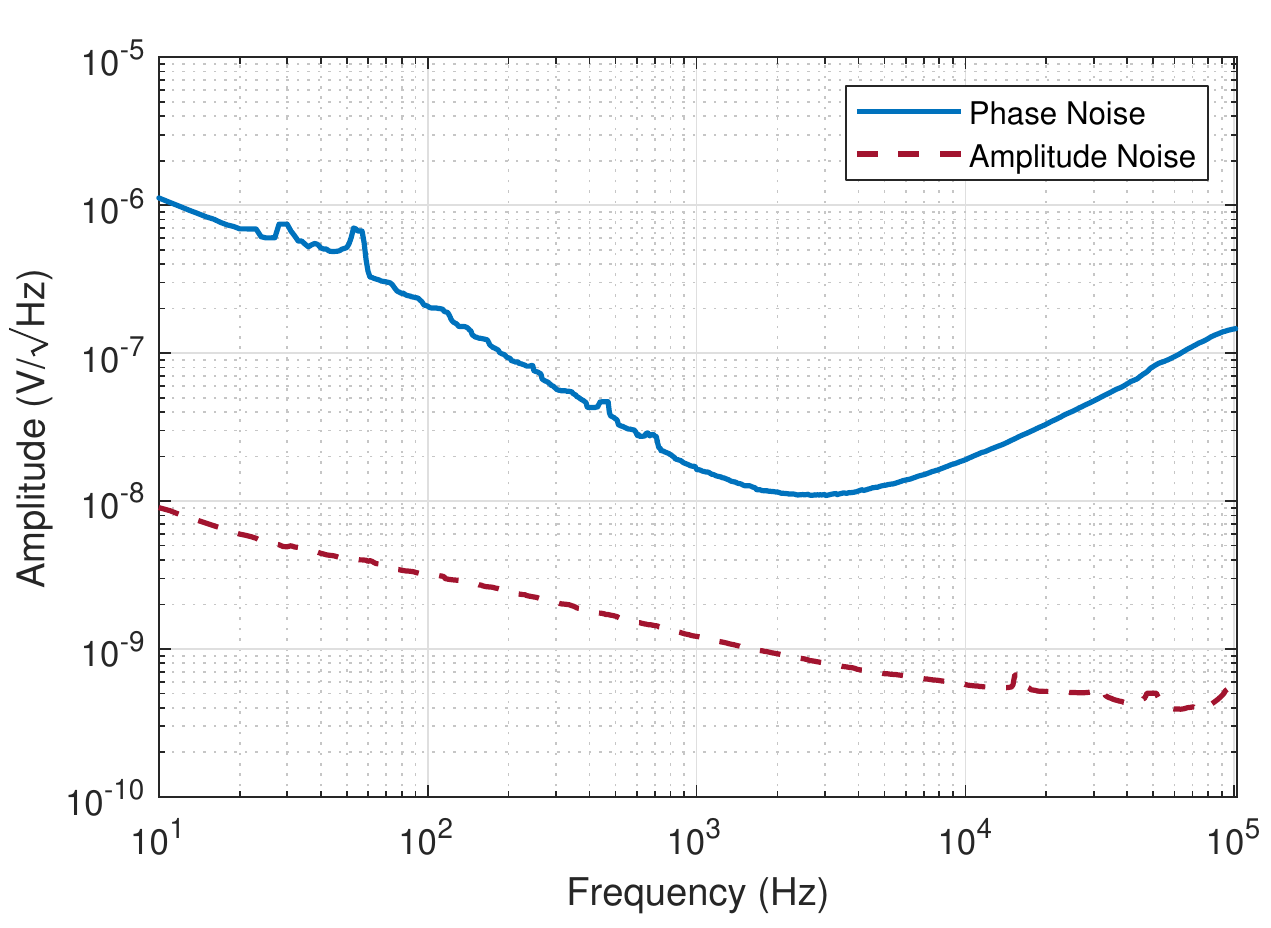}
\caption{\textbf{Composition of the total noise floor.} Using Eq. \eqref{eq:phasenoiseoutput}, the total noise floor is decomposed into a contribution from the MW source phase noise (\textcolor{matlabblue}{\rule[0.75mm]{3mm}{.25mm}}), and a contribution from the MW source amplitude noise (\textcolor{matlabred}{\rule[0.75mm]{1.1mm}{.25mm}~\rule[0.75mm]{1.1mm}{.25mm}}).
The phase noise contribution dominates amplitude noise at all offset frequencies analyzed.
}
\label{fig:am_vs_pn}
\end{figure}

By comparing $A(\omega)\Gamma_s(\omega)$ against $\Phi(\omega)\Gamma_p(\omega)$, the relative contributions of amplitude noise and phase noise to the noise floor are revealed. These two noise contributions are plotted in Fig.~\ref{fig:am_vs_pn} using the data from Fig.~\ref{fig:cavitysweeps}a and Fig.~\ref{fig:noiseshape}a. This reveals that phase noise is larger than amplitude noise by at least an order of magnitude for all offset frequencies of interest. Therefore, the contribution of the MW source to the total noise floor is dominated by phase noise.

\subsection{Thermal and phase noise limitations}
\label{supp:thermalnoise}
While the experimentally projected sensitivity is limited by phase noise of the MW source, this limit can be circumvented in the future with improved electronics. Thermal noise in the MW signal chain, however, is fundamental and will limit the attainable sensitivity at a given temperature. Given the measured slope $M_{\mathrm{max}}$ in Fig.~\ref{fig:cavitysweeps}b, the thermal-noise-limited sensitivity can be computed as
\begin{equation}
\label{eqn:thermallimit}
\eta_{\mathrm{th}} = \frac{G \sqrt{k_B TR}}{F M_{\mathrm{max}}}
\end{equation}
where $G$ is the total gain of the amplifier and mixer, $k_B$ is Boltzmann's constant, $T$ is the temperature, $R=50~\Omega$ is the termination impedance, and $M_{\mathrm{max}}$ 
denotes the maximum slope of the reflected RMS voltage under application of a sweeping bias magnetic field. $F$ is a constant of order unity that depends on the signal processing employed.
In practice, $F = \sqrt{2}$ because our signal is isolated to one channel of the IQ mixer.
With an optimized slope of $M_{\mathrm{max}}=2994$~$\mathrm{V}/\mathrm{T}$ and total gain $G = 21$~dB, Eq.~\eqref{eqn:thermallimit} yields a thermal-noise-limited sensitivity of $\eta_{\mathrm{th}}=1.1$~$\mathrm{pT}/\sqrt{\mathrm{Hz}}$.

For the sensitivity to be limited by thermal noise, the phase noise contributions must be less than the thermal noise contributions. We replace the cavity with a 50~$\Omega$ termination to measure the noise contribution due to thermal, amplifier, and digitizer noise, giving $e_{\mathrm{th}} = 13$~$\mathrm{nV}/\sqrt{\mathrm{Hz}}$ at 5~kHz offset.
The remaining noise contribution is from the MW signal source which is dominated by phase noise (see Supplemental Sec.~\ref{supp:noiseshape}).
Subtracting in quadrature from the total noise floor $e_n = 26$~$\mathrm{nV}/\sqrt{\mathrm{Hz}}$ yields the phase noise $e_p = 22$~$\mathrm{nV}/\sqrt{\mathrm{Hz}}$.
The phase noise of the signal generator output is directly measured by a Rhode \& Schwarz FSWP phase noise analyzer as $\Phi = -129.5$~dBc/Hz at 5~kHz offset.
For the sensitivity to be limited by thermal, amplifier, and digitizer noise, the MW source phase noise must therefore satisfy
\begin{equation}
\label{eq:phasenoisethermallimit}
\Phi_{\mathrm{th}} = \Phi \frac{e_{\mathrm{th}}}{e_p}\ell,
\end{equation}
where $0\leq \ell \leq 1$ determines how much lower the phase noise should be relative to the thermal noise. Taking $\ell=-6$~dB gives the phase noise requirement $\Phi_{\mathrm{th}} = -140$~dBc/Hz.

\begin{table*}[h]
\centering
\caption{Partial list of constants and parameters} 
\centering 
\begin{tabular}{l l c c } 
\hline\hline   
Name & Symbol & Approx. value & Units  \\
\hline   
Ruby zero-field splitting parameter & $D$ & $-2\pi\times 11.5\times 10^9/2$ & rad/s \\
Axial g-factor & $g_{||}$ & $\approx 2$ & unitless \\
Transverse g-factor & $g_{\perp}$ & $\approx 2$ & unitless \\
Bohr magneton & $\mu_B$ & $9.274\times10^{-24}$ & J/T \\
Electron gyromagnetic ratio & $\gamma$ & $2\pi\times 28\times 10^{9}$ & Hz/T   \\
Vacuum permeability & $\mu_0$ & $1.257\times 10^{-16}$ & H/m \\
Boltzmann constant & $k_B$ & $1.381\times 10^{-23}$ & J/K \\
System temperature & $T$ & 293 & K \\
Effective polarization & $\mathcal{P}$ & $4.7\times 10^{-4}$ & unitless   \\
Ruby resonator modal field volume & $V_{\mathrm{cav}}$ & 52.2 & mm$^3$  \\
Signal chain gain & $G$ & 21 & dB   \\
Reflected RMS voltage & $V_{\mathrm{RMS}}$ & & V    \\
Number of polarized spins & $N$ & $3.5\times 10^{14}$ & unitless    \\
Total number of spins & $N_{\mathrm{tot}}$ & $8\times 10^{17}$ & unitless    \\
Number of MW photons in cavity & $n_{\mathrm{cav}}$ & $2\times 10^{14}$ & unitless    \\
Cavity resonance frequency & $\omega_c$ & $2\pi\times 11.4\times 10^9$ & rad/s\\
Spin resonance frequency & $\omega_s$ & $\approx2\pi\times 11.4\times 10^9$ & rad/s\\
MW drive frequency & $\omega_d$ & $\approx2\pi\times 11.4\times 10^9$ & rad/s\\
Single-spin photon coupling & $g_s$ & $2\pi\times0.2$ & rad/s \\
Effective photon coupling & $g_{\mathrm{eff}}$ & $2\pi\times3.5\times 10^6$ & rad/s \\
Intrinsic linewidth & $\kappa_{c0}$ & $2\pi\times 320\times 10^3$ & rad/s\\
Input coupling rate & $\kappa_{c1}$ & $2\pi\times 340\times 10^3$ & rad/s \\
Loaded linewidth & $\kappa_c = \kappa_{c0}+\kappa_{c1}$ & $2\pi\times660\times 10^3$ & rad/s \\
Spin resonance linewidth & $\kappa_s$ & $2\pi\times 42\times 10^6$ & rad/s \\
Thermal polarization rate & $\kappa_{\mathrm{th}}$ & $2\pi\times 120\times 10^3$ & rad/s    \\
Loaded quality factor & $Q_L$ & 17,000 & unitless \\
Unloaded quality factor & $Q_0$ & 35,000 & unitless \\
Cooperativity & $\xi$ & 1.8 & unitless  \\
Bias magnetic field & $B_0$ & 3.1 & mT \\
Test magnetic field RMS amplitude & $B_\mathrm{test}^\mathrm{RMS}$ & 242 & nT \\
RMS voltage noise floor & $e_n$ & 26 & nV/$\sqrt{\mathrm{Hz}}$  \\
RMS phase noise & $e_p$ & 22 & nV/$\sqrt{Hz}$  \\
RMS thermal, amplifier, digitizer noise & $e_{\mathrm{th}}$ & 13 & nV/$\sqrt{Hz}$  \\
Test sensing magnetic field frequency & $f_m$ & 10 & Hz \\
RMS voltage peak at $f_m$ & $V_m$ & 0.65 & mV   \\
Maximum $V_{\mathrm{RMS}}$ slope while sweeping $B_0$ & $M_{\mathrm{max}} = \left[\frac{\mathrm{d}V_{\mathrm{RMS}}}{\mathrm{d}B_0}\right]_{\mathrm{max}}$ & 2994 & V/T \\
IQ signal processing gain & $F$ & $\sqrt{2}$ & unitless \\
Resistance & $R$ & 50 & $\Omega$  \\
Projected magnetic sensitivity & $\eta$ & 9.7 & pT/$\sqrt{\text{Hz}}$ \\
Thermal-noise-limited sensitivity & $\eta_{\mathrm{th}}$ & 1.1 & pT$/\sqrt{\mathrm{Hz}}$    \\
\hline 
\end{tabular}\label{tab:varnames}
\end{table*}

\cleardoublepage

\end{document}